\input harvmac

\def \sm {$\s$-model }

 \def \four{{\textstyle{1\ov 4}}}
\def \ov {\over}

\def \k {\kappa}
\def  \cN {{\cal N}}
\def  \rN {{\rm N}}

\def \S {{\cal S}}
\def \ss {{\cal S}}

\def \pa { \partial}
\def \a {\alpha}
\def \E {{\cal E}}
\def \b {\beta}
\def \g {\gamma}

\def \l {\lambda}

\def \m {\mu}
\def \n {\nu}

\def \s {\sigma}

\def \r {\rho}
\def \t {\theta}

\def \p {\phi}

\def \vp {\varphi}

\def \frac#1#2{{ #1 \over #2}}
\def \lr { \lref}

\def \lr{\lref}

\def \rf {\refs}

\def \adss {$AdS_5 \times S^5\ $}

\def \s { \sigma }

\def \vp {\varphi}

\def \p {\phi}
 
 \def \a { \alpha}
\def \r {\rho}

\def \DD {{\cal D}}

\def \zz {{\rm z}}

\def \DD {{\rm D}}

\def \t {\theta}

\def \del{\partial}
\def \m {\mu }
\def \n {\nu }
\def \ha { { 1 \over 2}}
\def \cN {{\cal N}}

\def \g {\gamma}

\def \k {\kappa}
\def \l {\lambda}

\def \b{\beta}

\def \ha {{1 \over 2}}

\def \ov {\over}
 \def \Om {\Omega}

\def \w {w}
\def \sss {{\textstyle{ 2 \ss \ov \sqrt{1 + \nu^2}}}}

\def \om {\omega}
\def \w  {\omega}

\def \sql {{\sqrt{\l}}}

\def \D {{\rm D}}
\def \E {{\cal E}}

\def \bd {{\bar \del}}

\def \E  {{\cal E}}
\def \D {{\rm D}}

\def \PP {{\cal P}}
\def \DD {{\cal D}} 

\def \t {\tau}
\def \rf {\refs}

\def \ss { s}

\def \rN {{\rm N}}

\def \bX {{\bar X}}
\def \D {\Delta}
\def \V {{\rm V}}

\def \tt {{\rm t}}

 \def \D {\Delta}
\def \ov {\over}
\def \adss {$AdS_5 \times S^5\ $}
\def \sm {sigma-model \ }
 \def \D { \Delta}

\def \N {{\rm N}}

\def \d {\del}
\def \bd { \bar \del}
\def \sql {{\sqrt \lambda}}
\def \el {\ell}
\def \ge {{\rm g}}
\def \bt {\ge }
\def \adn {$AdS_{\N-1}$}

\lref\hig{
A.~Higuchi,
``Massive Symmetric Tensor Field In Curved Space-Time,''
Class.\ Quant.\ Grav.\  {\bf 6}, 397 (1989).
``Symmetric tensor spherical harmonics on the N-sphere and their application to the de Sitter group SO(N,1)'', 
J. Math. Phys. 28 (1987) 1553.}

\lref \gkp{S.~S.~Gubser, I.~R.~Klebanov and A.~M.~Polyakov,
``A semi-classical limit of the gauge/string correspondence,''
Nucl.\ Phys.\ B {\bf 636}, 99 (2002)
[hep-th/0204051].
}

\lref\dev{
H.~J.~de Vega and I.~L.~Egusquiza,
``Planetoid String Solutions in 3 + 1 Axisymmetric Spacetimes,''
Phys.\ Rev.\ D {\bf 54}, 7513 (1996)
[hep-th/9607056].
}
\lr \bmn { D.~Berenstein, J.~Maldacena and H.~Nastase,
``Strings in flat space and pp waves from N = 4 super Yang Mills,''
JHEP {\bf 0204}, 013 (2002)
[hep-th/0202021].
}
\lref\mz{ J.~A.~Minahan and K.~Zarembo,
``The Bethe-ansatz for N = 4 super Yang-Mills,''
hep-th/0212208.
}

\lref\weg{
F. Wegner, ``Anomalous dimensions of high-gradient 
operators in the n-vector model in $2 + \epsilon$
dimensions'', 
Z. Phys. B78 (1990) 33.}

\lref\bre{
E.~Brezin and S.~Hikami,
``Fancy and facts in the (d-2) expansion of non-linear sigma models,''
cond-mat/9612016.
}

\lr \mina{
J.~A.~Minahan,
``Circular semiclassical string solutions on \adss,''
Nucl.\ Phys.\ B {\bf 648}, 203 (2003)
[arXiv:hep-th/0209047].
}

\lref\cast{
G.~E.~Castilla and S.~Chakravarty,
``Is the phase transition in the Heisenberg model described by the  (2+epsilon) expansion of the non-linear sigma-model?,''
Nucl.\ Phys.\ B {\bf 485}, 613 (1997)
[cond-mat/9605088].
}
\lref\pol{A.~M.~Polyakov,
``Gauge fields and space-time,''
Int.\ J.\ Mod.\ Phys.\ A {\bf 17S1}, 119 (2002)
[hep-th/0110196].
} 

\lr \mtt {
R.~R.~Metsaev and A.~A.~Tseytlin,
``Superstring action in \adss: kappa-symmetry light cone gauge,''
Phys.\ Rev.\ D {\bf 63}, 046002 (2001)
[hep-th/0007036].
R.~R.~Metsaev, C.~B.~Thorn and A.~A.~Tseytlin,
``Light-cone superstring in AdS space-time,''
Nucl.\ Phys.\ B {\bf 596}, 151 (2001)
[hep-th/0009171].
}

\lr \frt {
S.~Frolov and A.~A.~Tseytlin,
``Semiclassical quantization of rotating superstring in \adss,''
JHEP {\bf 0206}, 007 (2002)
[hep-th/0204226].
}

\lr \poll{A. Polyakov, talk at Strings 2002,
www.damtp.cam.ac.uk/strings02/avt/polyakov/ }

\lr \corr {
D.~Berenstein, R.~Corrado, W.~Fischler and J.~M.~Maldacena,
``The operator product expansion for Wilson loops and surfaces in the  large N limit,''
Phys.\ Rev.\ D {\bf 59}, 105023 (1999)
[hep-th/9809188].
}

\lr \POL {  A.~M.~Polyakov,
``String theory and quark confinement,''
Nucl.\ Phys.\ Proc.\ Suppl.\  {\bf 68}, 1 (1998)
[hep-th/9711002].
}

\lr \mald { J.~Maldacena,
``Wilson loops in large N field theories,''
Phys.\ Rev.\ Lett.\  {\bf 80}, 4859 (1998)
[hep-th/9803002].
}

\lr \calg{C.~G.~Callan and Z.~Gan,
``Vertex Operators In Background Fields,''
Nucl.\ Phys.\ B {\bf 272}, 647 (1986).
}

\lr \mets { R.~R.~Metsaev and A.~A.~Tseytlin,
``Type IIB superstring action in AdS(5) x S(5) background,''
Nucl.\ Phys.\ B {\bf 533}, 109 (1998)
[hep-th/9805028].
}
\lr \wein{
S.~Weinberg,
``Coupling Constants And Vertex Functions In String Theories,''
Phys.\ Lett.\ B {\bf 156}, 309 (1985).
}

\lr \yon {
S.~Dobashi, H.~Shimada and T.~Yoneya,
``Holographic reformulation of string theory on \adss background
 in the PP-wave limit,''
hep-th/0209251.
}
\lr \GKP{S.~S.~Gubser, I.~R.~Klebanov and A.~M.~Polyakov,
``Gauge theory correlators from non-critical string theory,''
Phys.\ Lett.\ B {\bf 428}, 105 (1998)
[hep-th/9802109].
}
\lr \WI{
E.~Witten,
``Anti-de Sitter space and holography,''
Adv.\ Theor.\ Math.\ Phys.\  {\bf 2}, 253 (1998)
[hep-th/9802150].
}

\lr \zar{K.~Zarembo,
``Open string fluctuations in \adss and operators with large R
 charge,''
hep-th/0209095.
}

\lr \ts { A.~A.~Tseytlin,
``Semiclassical quantization of superstrings: \adss and beyond,''
hep-th/0209116.
}

\lr \kra{V.~E.~Kravtsov, I.~V.~Lerner and V.~I.~Yudson,
``Anomalous Dimensions Of High Gradient Operators In The Extended Nonlinear Sigma Model And Distribution Of Mesoscopic Fluctuations,''
Phys.\ Lett.\ A {\bf 134}, 245 (1989).
}

\lr \buch{
I.~L.~Buchbinder, E.~S.~Fradkin, S.~L.~Lyakhovich and
V.~D.~Pershin,
``Higher spins dynamics in the closed string theory,''
Phys.\ Lett.\ B {\bf 304}, 239 (1993)
[arXiv:hep-th/9304131].
V.~D.~Pershin,
``Interaction of higher spin massive fields with gravity in string
 theory,''
arXiv:hep-th/0009248.
}

\lr \fri{
D.~H.~Friedan,
``Nonlinear Models In Two + Epsilon Dimensions,''
Annals Phys.\  {\bf 163}, 318 (1985).
}

\lr \rey{
S.~J.~Rey and J.~Yee,
``Macroscopic strings as heavy quarks in large N gauge
theory and  anti-de Sitter supergravity,''
Eur.\ Phys.\ J.\ C {\bf 22}, 379 (2001)
[hep-th/9803001].
S.~J.~Rey, S.~Theisen and J.~T.~Yee,
``Wilson-Polyakov loop at finite temperature in large N
gauge theory and  anti-de Sitter supergravity,''
Nucl.\ Phys.\ B {\bf 527}, 171 (1998)
[arXiv:hep-th/9803135].
}

\lr \SZ{
G.~W.~Semenoff and K.~Zarembo,
Nucl.\ Phys.\ Proc.\ Suppl.\  {\bf 108}, 106 (2002)
[arXiv:hep-th/0202156].
}

\def \N {{\rm N}}
\def \bx {{\bar N_x}}
\def \x {{\rm x}}

\def \bnx{ \bar n_x}

\Title{\vbox
{\baselineskip 10pt
{\hbox{
 }
}}}
{\vbox{\vskip -30 true pt
\centerline {On semiclassical  approximation }
\centerline { and spinning  string vertex operators in $AdS_5 \times S^5$ }
\medskip
\medskip
\vskip4pt }}
\vskip -20 true pt
 \centerline{
  A.A. Tseytlin
  $^{a,b,}$\footnote{$^{*}$}
 {Also at
 Lebedev Physics Institute, Moscow.}
 }
 \smallskip\smallskip
 \centerline{ $^a$ \it  Blackett Laboratory,
  Imperial College,
  London,  SW7 2BZ, U.K.}

 \centerline{ $^b$ \it  Department of Physics,
  The Ohio State University,
  Columbus, OH 43210, USA}

\bigskip\bigskip
\centerline {\bf Abstract}
\baselineskip12pt
\noindent
\medskip
Following earlier work  by Polyakov and Gubser, Klebanov and Polyakov,
 we attempt to clarify  the structure of  vertex 
operators representing  particular string  states 
which  have large (``semiclassical'') values of 
AdS energy or 4-d dimension $E=\D$ 
and angular momentum $J$ in $S^5$ or spin  $S$ in $AdS_5$.
We comment on  the meaning of semiclassical limit in the context 
of $\a'\sim {1 \ov \sql}$ perturbative expansion for the 
2-d anomalous dimensions of the corresponding vertex operators.
We consider in  detail the leading-order 1-loop 
renormalization of these operators in $AdS_5 \times S^5$ 
 sigma model 
(ignoring fermionic contributions). We  find 
 more examples of operators  (in addition to  the one 
in hep-th/0110196) for which the  1-loop anomalous dimension 
can be made small by tuning  quantum numbers. 
We also  comment on  a possibility of deriving the 
semiclassical relation between $\D$ and $J$ or $S$  
from the marginality condition for the vertex operators, 
using  a stationary phase approximation in  the path integral 
expression for their 2-point correlator on a complex plane.

\Date{April 2003}

\noblackbox
\baselineskip 16pt plus 2pt minus 2pt

\newsec{Introduction}

One of the 
important  issues  in gauge theory -- string theory duality is 
to  understand
 the precise  formulation
of the relation between observables on the two sides of 
the duality
beyond the supergravity (chiral primary) mode sector.
Progress in that direction was initiated
in \rf{\pol,\bmn,\gkp}.
A  natural   expectation  \pol\ 
 is that each on-shell string state  or
 marginal string vertex operator in $AdS_5 \times S^5$ string theory 
should be associated with a local 
gauge-invariant  operator  in $\cal N$=4 SYM theory
with definite   conformal dimension $\D= \D(Q,\l)$ 
($Q$ are global charges and $\l$ is the 
`t Hooft coupling).\foot{A related 
 formulation or manifestation of duality
is the equality of the gauge theory and string theory 
expectation values for a  Wilson loop
\rf{\POL,\mald,\rey,\corr,\SZ}.}
Then  the original suggestion of \rf{\GKP,\WI} implies   that
one is supposed  to relate  correlators of local
gauge-invariant conformal operators  on the gauge theory
 side  with  elements of string ``S-matrix'' in AdS, i.e. with 
correlators of the appropriate local 
string vertex operators  
(having the same  quantum numbers and  corresponding to a 
specific  choice of  boundary conditions).


Our aim here is to try to shed light on 
 the structure 
of  string vertex operators  corresponding  to particular string states 
in \adss  with  large 
values of $AdS_5$ energy $E$  in global coordinates 
(equal to  4-d  dimension $\D$) 
 and   global charges $Q$  (e.g., angular momentum $J$ in $S^5$ 
 or  spin $S$ in $AdS_5$).
 We shall be guided by  the previous work in this direction 
in \rf{\pol,\poll} (see also \zar)  and  by 
the predictions  \gkp\ 
for $ \D= \D(Q)$ for the associated 
semiclassical string states.

In \gkp\ it was suggested  that  a spinning string state 
represented by  the classical solution for a long rotating string
in AdS  \dev\  may  be related,  in the limit of large spin $S$, 
to the minimal twist    operators
on the gauge theory side. An evidence in favour of this proposal 
is the scaling of the dimension  
 as 
 \eqn\sio{ \D=  S + c_1 \ln S +.... }
   for large $S$
in the two  limits ($\l \gg 1 $ and $\l \ll 1$)
 of the duality,  as well as the
absence \rf{\frt,\ts} of higher-order $(\ln S)^n$ corrections
on the string side to all orders in the 
inverse string tension (${1 \ov \sql}= { \a' \ov R^2} $).
Similar behaviour $\D= J+c_2 + ...$  was found \gkp\ also 
(in the limit of large $J$)
  for a stretched string rotating in $S^5$. 

In flat space
one may  represent a spinning 
string state on the  leading Regge trajectory
either  by a local vertex operator $ e^{-iEt} (\del X 
 \bar  \del X)^{S/2} $, 
\ $ X= x_1 + i x_2$ 
 (or as a Fock space state 
  $( a^\dagger_1  a_{-1} ^\dagger)^{S/2}  |0,E>$)
or, if the spin $S$ is of   order of string tension, 
by a semiclassical  state
described  by a ``solitonic'' rotating string
solution (in Fock space this is a 
coherent state exp$( \sqrt S a^\dagger_1
+ \sqrt S a_{-1} ^\dagger) |0,E>$ with respect to  the
elementary  oscillator vacuum).

By analogy,  one should expect that the same $\D= \D(S)$ relation 
as found for  the ``solitonic''  rotating string solution in AdS 
should  be found also  for 
an  ``elementary''  string state  created by a
 local vertex operator $V_{S}$
with the same  energy $E=\D$  and spin $S$ 
(and having the  correct flat-space limit  representing 
state on a leading Regge trajectory).
The solutions in \gkp\ 
should be associated with 
 closed string states on the  2-d cylinder.
 Via the standard  conformal mapping,  
they may be expected to  be in correspondence 
with states created by the two vertex operators inserted 
 on a complex plane 
with the  ``mass-shell''  condition $\D=\D(S)$ 
 following  from the 
marginality condition of the vertex operators.

Identifying  string   vertex operators  
 in \adss superstring theory is, in general, 
  a complicated 
task. First, they  may  have a  non-trivial
 dependence on fermions.
In addition, 
vertex operators in curved space may 
 in general depend on $\a'$.
However, one may hope that in  the semiclassical approximation
considered in \gkp, i.e. ($E=\D$) 
\eqn\cla{
  {\D\ov \sql}=\k= {\rm fixed}
\ ,  \ \ \ \ \ \ \ \ {Q\ov \sql}=q=
{\rm fixed}
\ , \ \ \ \ \ \ {1 \ov \sql} \ll 1 \ ,  \ \ \ \ \ Q= J \ {\rm or} \ S \ , }
 where  
$\sql$ is related to the effective string tension  $T= { R^2 \ov 2 \pi \a'}
 = {\sql \ov 2 \pi}$,  there may be some simplifications.  
In particular, 
given that the spinning classical solution in \gkp\  involved 
only the bosonic AdS 
 part of the \adss string theory, 
the result for $\D=\D(S)$  may be universal,  i.e. 
it may not be sensitive to the  fermionic part of 
the vertex operator
and  details of  $S^5$ factor. 
With this expectation 
 we shall concentrate below only on  bosonic parts of the 
corresponding operators.

Instead of computing the form of the marginality condition 
for  a spinning string   vertex operator 
directly in perturbation theory in $ {\a'\ov R^2}  = { 1\ov \sql}$ 
one may hope that in the  limit of large $\D$ and $S$ 
the relation $\D(S)$ may be derived  by using semiclassical approximation 
in the string path integral on the 2-sphere (complex plane) 
with two  insertions  of the  vertex operators  $V_{S}$.
This  idea works indeed in the case of a point-like string mode
(supergravity state)  
with large $S^5$ orbital momentum  $J$ \rf{\gkp,\poll}.
One could then try  to 
follow the same logic  in the spinning  case, i.e. start with 
the 2-point function  of unintegrated
operators on a 2-sphere,   
 compute the dependence on the 2-d distance  $|\xi_1-\xi_2|$
in the semiclassical  approximation
and show that insisting that the 2-point 
correlator  scales as  $|\xi_1-\xi_2|^{-4}$
implies the same relation
$\D=\D(S)$ as found  in \gkp.
 Such computation  can indeed be 
 done for a spinning string state  in flat space 
  where, as we shall  explain below,  
it  leads indeed to the Regge trajectory relation 
$\D\equiv E = \sqrt { {2 \ov \a'} S}$.

Indeed, 
since in the semiclassical limit \cla\  $S$ scales as string tension,
 one expects the 2-point correlator to be
saturated by a classical trajectory  which should be
closely related  to the rotating  string solution  in \gkp.\foot{
The discussion   in \gkp\  applied to 
string theory in Minkowski-signature  AdS space
defined on a 2-d  Minkowski-signature cylinder.
This describes propagation of a particular closed  string
mode in real time.  The vertex-operator 2-point
computation  done on Euclidean 2-sphere
may be mapped onto Euclidean 2-cylinder with the vertex
operators specifying a particular  string state
propagating on the cylinder.
It is natural to expect that 
the relevant semiclassical trajectory should then be a (complex
and conformally transformed)
analog of the one in \gkp. As already apparent  in
the $\D=J$ case \refs{\poll,\zar}, one should not
 attribute a special meaning to the
complex nature of the semiclassical trajectory
saturating path integral (cf. \yon).
 An imaginary  nature of it may be related  to 
 external
sources one puts in to specify the required
 boundary/initial  conditions.
Like in the case of a euclidean gaussian path integral
 with imaginary sources   the result is an analytic function of 
  $J$ so that  one can make analytic continuations.}
Presumably, the role of the two  vertex operators  is only  to
insert proper boundary conditions in mapping the 2-sphere back 
to the cylinder, while their 
detailed pre-exponential form should not be  that
important.
Unfortunately, we were  not be able  to find   a  precise 
implementation of this idea  in the $AdS_5$ (or $S^5$) 
spinning string  case.  


 One may wonder  how the expression for $\D(S)$ found  in semiclassical 
approximation may be related  to the direct computation 
of the 2-d anomalous dimension of the corresponding vertex operator  in 
sigma model perturbation theory in $1 \ov \sql$.  
In general, an eigenvalue of the anomalous dimension 
matrix   
for the operator  at level (i.e. with  number of 2-d  derivatives) 
equal to $Q$  and with quantum numbers $\D$ and $Q$  is
 expected to have the structure
$$
\g = 2 - Q  + { 1 \ov \sql} ( a_1 \D^2 + a_2  Q^2 +  a_3 \D + a_4 Q)  $$
\eqn\ssu{
+ \ 
{ 1 \ov (\sql)^2 } ( b_1  \D^3 + b_2  Q^3 +  b_3 \D^2 + b_4 Q^2 +  b_5 \D + b_6 Q)
+ ... +  
{ 1 \ov (\sql)^n } ( c_1  \D^{n+1}  + c_2  Q^{n+1}  +  ...) + ... \ . }
The marginality condition 
\eqn\maar{ \g=0   }
should then produce a complicated   relation 
$\D = \D( Q, \sql)$. 
Equivalent relation should follow from solving the generalized 
Klein-Gordon type   equation  $\hat \g f = (2-Q + \ha \a'  \nabla^2 
+ ... )f= 0 $
 for the corresponding wave function $f$, with the operator  
  $\hat \g  $ 
representing  the functional form  of the anomalous
 dimension operator  \rf{\fri,\calg}. 
 
While determining the detailed form of higher order corrections 
in \ssu\  is in  general a very complicated problem depending   
 on  details 
of both the superstring action and  particular vertex  operator, 
 a  semiclassical path integral computation of the 
 anomalous dimension 
would  bypass this problem, effectively 
summing up leading terms in \ssu\ from each order  in $1 \ov \sql$.
That all orders in $\a'$ should be contributing  is obvious  
from the form of the semiclassical relation  found in \gkp\ 
in the limit \cla: $ \D = \sql\ \k (q) = \sql\
 \k ({ Q \ov \sql})$. 
Indeed, assuming that $\D$  and $Q$ scale with $\l$ as in  
  \cla\  and thus keeping only the leading terms 
in $\D$ and $Q$ at each order in $1 \ov \sql$ one finds from  \ssu\ 
\eqn\dqg{
\g = 2 - \sql f(\k,q) +  h(\k,q) + O({1 \ov \sql})  \ , }
$$ f= q +  a_1 \k^2 + a_2  q^2  +  b_1  \k^3 + b_2  q^3  + ...  \ , \ \ \ \ \ 
 h= a_3 \k + a_4 q + b_3 \k^2 + b_4 q^2 + ...  \ , $$
 so that the  leading-order 
 solution of $\g=0$ condition at large $\l$ should 
 be determined  by solving $ f(\k,q) =0$, which gives 
  $\k = \k_0 (q) + O({1 \ov \sql})$.
Remarkably, the 
 semiclassical argument should thus be effectively 
 determining  the whole function $f(\k,q)$ 
which  contains information coming from all higher-order
  sigma  model 
loop corrections to the anomalous dimension. 

One may expect that  some features of the semiclassical relation \sio\ for 
$\D(S)$  can be seen already in the  perturbative  expansion \ssu. 
For example,  one would be able to  deduce 
that $\D= S + ...$ at large 
$S$ provided  {\it each } of the  expansion  terms
 in \ssu\ would have the form 
$ \D^{n+1} - S^{n+1} + ...$. In particular,  
the one-loop  correction 
would  then  to  start with  $\D^2 - S^2+ ...$, similarly 
 to  what happens in the example (corresponding to 
  a scalar string 
state)  pointed out  in \pol. 
As  we shall discuss below, such ``difference of squares''
form of the leading 1-loop correction to the 2-d anomalous 
dimension is indeed  characteristic also 
to vertex operators representing  spinning  string states.
However, the  relative coefficient of the two terms
in the difference does not turn out to be exactly one. 
One possibility is that the precise value of this 1-loop coefficient 
depends in fcat  on the ignored  contribution of the fermionic terms 
(both in the string sigma model action 
and the vertex operator).
An alternative explanation is that one is first to sum all 
orders in $1\ov \sql$ to determine the function $f( {\D\ov \sql},
{ S\ov \sql})$ in \dqg\ and then  find $\D=S + ...$ 
only after solving $f=0$.



We shall start in section 2 with a review of the
simple  $S^5$ orbital  rotation case \rf{\bmn,\poll}.
We shall  then consider in section 3 
the semiclassical  path integral derivation of the $E\sim  \sqrt S $ relation
for spinning string vertex operator in flat space.

In section 4 we shall describe the
expected   structure of vertex operators 
corresponding to string rotating in $S^5$ and to string rotating in $AdS_5$.
We shall then sketch a possible generalization of 
the semiclassical argument in flat space to the case of the string 
spinning in $S^5$. 
 
 With the motivation to find the  precise structure 
 of the spinning string vertex operators in \adss in  section 5 we
  shall study in detail the 1-loop renormalisation of composite 
  operators in $O($N$)$ invariant bosonic sigma model. We shall 
 follow \pol\ and  apply the method of \weg. 
  We shall find that in general the vertex operators are given by 
  mixtures  
  of various possible operators of the same canonical dimension
  and spin, with 1-loop corrected  eigenvalues of the anomalous dimension 
  matrix  having  a ``difference of squares''  structure
  as in the example considered in \pol.
  Assuming  that higher loop corrections will  have similar structure, 
  this seems to be in  qualitative agreement 
   with the 
  semiclassical results of \gkp\  ($\D \approx S$ and $\D \approx J$)
   for the corresponding string states on a 2-d cylinder.

\newsec{Case of orbital angular momentum in  $S^5$: review}

\subsec{\bf Comments on scalar vertex operators in $AdS$} 

Let us briefly review  some  relevant points
regarding  the general structure
of  integrated and unintegrated vertex operators in AdS space 
\pol.
In any curved background  we can define  string vertex
operators  as perturbations of string sigma model action  that
solve the conformal invariance  (marginality) conditions. The latter 
are analogs of curved space Laplace equation with various
$\a'$ corrections. These equations are known explicitly in very few
cases (gauged WZW models, etc). As discussed above,
for certain types of string modes  the
 semiclassical approach
may allow  one to fix implicitly the form of the leading terms 
in  these equations,  i.e. to determine
their solutions  with proper boundary conditions
and thus  determine the  space-time dimensions of the associated
CFT operators.

Ignoring $\a'$-corrections, a scalar mode like dilaton
 should satisfy the KG equation  $( -  \nabla^2  + m^2)  T=0$ 
   in \adss.
 To define  vertex operators leading to 
an analog of ``S-matrix'' we need to  
 specify  appropriate  boundary conditions  \rf{\GKP,\WI}. 
In flat space  the relevant solutions are  
   plane waves with fixed momentum, i.e. vertex operators
 contain factors of $e^{ipx}$.
  In AdS,  while
 a 2-d  conformal  vertex   is  any  $ \hat V =
 \int d^2 \xi \  T(x(\xi)) $
 where  $T$  solves  the  KG equation,
 we  need  to choose a specific solution.
 The  required one is  ($k=1,...,d=4$) 
\eqn\sad{
T(\hat x) = \int d^4 x' \ K(\hat x, x')  T_0 ( x') \ , \ \ \ \ \ 
\hat  x= (z,  x) \ , \ \ \ \ \ 
ds^2 = { dz^2 + dx_\m dx_\m  \ov z^2}\ . }
Here   $T_0 (x) $ is any ``source''
 function  at the boundary of AdS,  
 $K$ is the  Dirichlet bulk to boundary propagator,
   \eqn\kek{  K = c(\D )  [{ z \ov z^2 + (x-x')^2}]^\D \ , \ \ \ \ \ \ 
K_{z\to 0} \to \delta^{(4)}  (x-x') \ ,  }
and $\D$ is determined from  the condition
 $0= \g= - \ha \a' m^2  + { 1 \ov 2 \sql} \D(\D-4) $.
 The  vertex operator that
 enters the expressions for string correlators  is the 
``unintegrated one'' --
the one  integrated over the 2-d world sheet
 but depending (instead of the usual  momentum) on a point
  at the  boundary of AdS:
 \eqn\opi{ \hat V(x)  =
 \int d^2 \xi\  \V (\xi) \ , \ \ \ \ \V=  K(\hat x(\xi),x)  \  ,\ \ \ \ \ \ \ 
\hat V(T_0) = \int d^4 x\ \hat V(x)\  T_0 (x) \  . }
 The AdS/CFT conjecture \rf{\GKP,\WI}
can  then   be formulated as the statement  that
 the  string generating functional as functional of $T_0(x)$
 is equal to the corresponding gauge-theory
generating functional,
 $< e^{ \int d^4 x \ T_0 (x) {\cal O}(x) } > $,
where ${\cal O}(x)$ is a local gauge-invariant
operator with appropriate quantum numbers and dimension.\foot{It does not 
seem to be known how
 to directly compute the  string 2-point
 function in this set-up. In compact 2-d CFT case the
 result for the 2-point correlator
 $<VV>$ is zero  due to division by the Mobius  group volume.
 In the non-compact
AdS case we are supposed to get a finite
result; this should be a consequence,  in particular, of 
the special 
 role of the boundary  terms in AdS. }
 Choosing $T_0$ to be a  delta-function gives us simply
 the correlators of unintegrated
vertex operators each  localised at a  given boundary point
and having specific dependence
on radial direction  $z$.

This (euclidean) AdS/CFT duality  prescription of the equality of the two generating
functionals \rf{\GKP,\WI}
 was not yet   tested at the full 
 string-theory (as opposed to supergravity)  level.
One motivation to study the structure of 
string-level  vertex operators  and  of 
semiclassical approach  in the string path integral
is that  this    may help to clarify 
the   duality relation.

\subsec{\bf Operators with orbital momentum $J$ and semiclassical approximation} 

Let us now  review  the  case of the simplest 
vertex operator  carrying large  orbital momentum  $J$ 
in $S^5$     and the semiclassical derivation of
its dimension. 

Consider  the  dilaton-type   vertex operator
corresponding to the ground-state (supergravity) 
level state. 
According to \opi, the 
 unintegrated operator localised at point 0  at the boundary can be
  written as 
 can be written as
(see \rf{\pol,\poll})
\eqn\tyu{
\V_J (\xi) \equiv \V_J (x(\xi),z(\xi),\vp(\xi)) = 
  C \  (N_+)^{-\D}\  (n_x)^J \  V^{(2)} \ ,  }
where $ V^{(2)} =  -\del N_M \bar \del N_M  + \del n_k  \bd n_k  + ...$
is a scalar operator that has the same form as the \adss sigma model 
Lagrangian
and is contributing to the  2-d dimension to make a $(1,1)$ operator.\foot{Dots stand for the  fermionic contributions
 ($\a'$-corrections are expected to be absent for a vertex operator 
representing a supergravity  state 
which is dual to a chiral primary operator, see also below). 
For the dilaton vertex operator
in flat $D$ dimensions defined on 2-sphere one is to add also the term 
$a R^{(2)} \ ,  \    a= \four \a' (D-2)$.}
In particular,   the Euclidean $AdS_5$ space and $S^5$  are  described by 
the two 6-vectors $(\m=0,1,2,3; \ k=1,...,6$)
\eqn\vecc{
N_M N_M \equiv  N_5^2  - N^2_\m - N^2_4  =1  \ , \ \ \ \ \ \ \ \ 
   \ n_k n_k =1 \ . } 
The linear combinations appearing in \tyu\ are\foot{Note that 
  $N_+$ scales under the dilation  isometry of AdS 
in the same way as the boundary coordinate
$x_\m$ ($z \to \l  z, \ x_\m \to \l  x_\m $). 
Let us also recall the relation between $N_M$, 
global $AdS_5$ coordinates ($t, \rho$, and $S^3$ angles
with metric $ds^2 = - 
 \cosh^2 \r \ dt^2 +  d\r^2 + \sinh^2\r \ d\Om_3$) 
and the Poincare coordinates:
$
N_0={x_0\ov z} = \cosh \r \ \sin t \ , \ \ \ \ \ 
N_5= { 1 \ov 2 z} ( 1 + z^2 - x^2_0 + x_i^2) =  \cosh \r
\ \cos t \ , $\ \ 
$
N_i= {x_i\ov z} =\hat  n_i \sinh \r  \ , \ \ \ \ \ \
N_4={ 1 \ov 2 z} ( -1 + z^2 - x^2_0 + x_i^2) =  \hat n_4
\sinh\r \ , $
so that $\tan t = { 2 x_0 \ov 1 + z^2  - x^2_0 + x_i^2 }$ and 
 $
z^{-1} = \cosh \r \ \cos t - \hat n_4 \sinh \r $.
Here the unit vector
 $\hat n_k$ ($k=1,2,3,4$), \     $\hat n^2_i + \hat n^2_4=1$, \
 parametrizes the 3-sphere: \ $d\hat n_k d\hat n_k = d\Omega_3$.
}
\eqn\nen{
N_+ \equiv  X_4 + X_5 = { 1 \ov z} ( z^2 + x_\m x_\m) \ , \ \ \ \ \ \ \
N_- = { 1 \ov z}  \ , 
}
and 
\eqn\zez{
n_x \equiv  n_1 + i n_2 = {z_1 + i z_2 \ov z} = e^{i  \vp }   \ ,  } 
which represents rotation in $(z_1,z_2)$ plane in 
$R^6$ or along the big circle of $S^5$.

To localise  such operator at the boundary  we need to
form a linear superposition integrating over $\xi$
\eqn\linr{
\hat V_J (x)= \int d^2 \xi \  \V_J(x(\xi) -x  ,z(\xi),\vp(\xi))
\ . }
This operator will have zero world-sheet dimension  and
the right   space-time dimension
to be dual to the corresponding gauge theory operator with dimension $\D$. 
 Then   for $z\to 0$ (cf. \kek)
\eqn\sdw{
\hat V_J(\x)
\to  \int d^2 \xi \  \delta^{(4)} (x(\xi) - x)
 \  ( n_z(\xi) )^J\   V^{(2)} (\xi) \ ,
}
i.e. we get indeed an operator localised  at one point at the boundary.\foot{
It has an arbitrary 4-momentum if we do the  Fourier transform
in $x$, i.e.  it is  ``on-shell'' in 5-d but
 ``off-shell'' in 4-d sense.}
The correlators of $\hat V_J$ computed using \adss
 string sigma model on
the 2-sphere  should give correlators of the dual gauge theory at the
boundary, with  $\hat V_{J}(x) \leftrightarrow {\cal O}_{J} (x)$.
On the basis of conformal symmetry we expect
\eqn\yuy{
< \hat V_J(x) \hat V_{{-J}}(x')>\  \sim\   |x - x'|^{-\D} \ .
}
To determine  $\D=\D(J)$  one may compute the anomalous dimension 
of \tyu \ using  sigma-model perturbation theory 
in $1\ov \sql$. One can show that at least in the 1-loop 
approximation (and ignoring all fermionic contributions)\foot{Contrary
to naive expectation, 
fermions may in principle contribute to 
renormalization of composite operators even in the 1-loop approximation:
the superstring action contains the RR  5-form coupling term \mets\ 
$ \sim \bar \theta  \theta  \del x , $  and 
the fermions in it  may be ``paired'' with the fermions in the 
$\del x +  \bar \theta  \theta $ factors in the  vertex operator.}
the operator \tyu\  is an eigen-operator of the 
anomalous dimension matrix  and so 
 the resulting marginality condition 
 takes  the form  \ssu\ 
(see \rf{\pol,\gkp}\ and below) 
\eqn\margi{
0=\g = 2 - 2  + { 1 \ov 2\sql} [ \D(\D-4) - J(J+4) ]  + O({ 1 \ov (\sql)^2 })
\ . }
This implies  $ \D= J +...$  for large $J$.

In the case of the 
large ``classical'' values of $\D$ and $J$ in \cla\ the
 same  relation can be  obtained 
using  a semiclassical approximation
in the path integral for the 2-point function of the unintegrated 
vertex operators \poll.
As was mentioned above in  \ssu,\dqg, this semiclassical approach 
in general goes beyond the perturbative 1-loop result.  


Using the global AdS coordinates in the string action, 
rotating  the global AdS  time $t$ to the Euclidean one $t_e=it$
(so that $N_+ \sim  e^{t_e}$) 
  and expanding near point
where $\r=0$ with all  angles except $\vp$ in \zez\  being  trivial,
  we get for  the relevant  part of the Euclidean string  action  on
the 2-sphere
\eqn\uii{
I=  {\sql \ov 4 \pi }\int d^2 \xi \ [ ( \del_a t_e)^2 ( 1 + ...) 
  + (\del_a \vp)^2 (1 + ...)   + ... ]
\ . }
The  aim is to show that demanding  that the two-point function 
has canonical 2-d dimension 
\eqn\ess{
 < \V_{J} (\xi_1) \V_{-J} (\xi_2) >
= \int [dt_e \ d\vp...]\  e^{-I[t_e,\vp,...]}
 \   \V_{J} (\xi_1) \V_{-J} (\xi_2)
\ \   \sim \  |\xi_1 - \xi_2|^{-4}
}
implies
that  $\D= J $.
The insertions of the vertex operators provide  effective source
terms to the above action:
\eqn\kokl{
 e^{-\D [ t_e(\xi_1) - t_e(\xi_2) ]   + i J [ \vp (\xi_1) - \vp (\xi_2) ]}  \ .   }
When $\D$ and $J$ have  large ``classical'' \cla\ 
 values, the action and the ``source'' terms 
are of the same order, and the path integral is saturated by 
the stationary point  trajectory. 
 Then (as in the usual case of a gaussian integral)
 we get a complex \poll\ semiclassical trajectory  for the angle $\vp$
($\vp =  {J\ov \sql}  ( \ln |\xi - \xi_1| -  \ln |\xi - \xi_2|)$).  
The final semiclassical  result for the  correlator \ess\ is then 
 proportional to
 \eqn\zaz{
 < \V_{J} (\xi_1) \V_{-J} (\xi_2) >  \ \
\sim \ \
|\xi_1 - \xi_2|^{2\g  - 4 } \ ,
\ \ \ \ \ \ \ \ \ \g = {1 \ov 2 \sql} (\D^2 - J^2) \ .  }
 Here   -4 comes  (as in  the flat space case) 
from the contribution of  $<V^{(2)} V^{(2)}>$.
Demanding the marginality of the vertex operators, i.e. $\g=0$, 
 we are thus 
led \poll\ to the relation 
$$\D=J  \ . $$ 
To relate this derivation to 
the classical solution of string theory defined
 on a  2-d
 cylinder one  may take $\xi_1=0$, map the point $\xi_2$ to $\infty$
  and 
then  map the complex $\zz$-plane with two punctures into the  Euclidean 
cylinder  by setting (see also section  3)   
\eqn\euc{
\zz= e^{\rm w}\ ,  \ \ \ \ \ \    {\rm w} = \tau_e + i \sigma    \ , \ \ \ \ 
\ \ \  \tau_e = i \tau . } 
 After we  rotate back to the Minkowski signature
the semiclassical trajectory will become  the same
as the geodesic in \bmn\ (i.e.  $t = \k \tau, \ \ \vp = \k \tau,  \  
 E= \D = \sql\ \k  $).

\newsec{\bf Semiclassical derivation of dimension 
of  spinning string   vertex operator  in flat space}
We would like in principle 
 to apply  a similar  semiclassical argument to the case 
of  the spinning  string states.
Though this  turns out to be rather hard to achieve 
in practice in \adss case, 
 as we will describe 
in this section  
   the  semiclassical derivation 
of the relation between energy and spin 
 works indeed in  the flat space case. 

The vertex operator  that describes  a (bosonic) 
 string state on the leading Regge trajectory 
with spin  $S$   and energy $E$  in flat space 
 can be represented as follows 
\eqn\fle{
\V_S(\xi) = e^{ -i E t } \big( \pa X\bar{\pa} X \big)^{S/2}\ ,   }
\eqn\lop{ X = x_1+ix_2,\ \ \ \ \ \   \bar{X}=x_1-ix_2 \ . }
For any  $S$ the  marginality  condition  implies that 
\eqn\maar{
\g = 2 - S  - \ha \a'   E^2 =0 \ , \ \ \ \ \ \ \ {\rm i.e.}
 \ \ \ \ \ \  \a'  E^2  = 2 (S-2)  \ . } 
In flat space  the  spin factor provides only a contribution to 
canonical dimension, while the anomalous  dimension comes from the
1-loop renormalisation of the  exponential (energy) factor. 

Let us show that for large (classical) value of $S$ 
the flat  Regge trajectory relation \maar\  can be found by a 
semiclassical 
path integral argument similar to the one in the previous section.
The  aim is to compute
\eqn\vvvv{
A_2 (\xi_1-\xi_2)\equiv  <\V_S(\xi_1) \V_{-S}(\xi_2)>\ 
, \ \ \ \ \ \ \ \ \ \  \V_{-S} \equiv 
 \V_S ( X \to \bar X) 
\  }
 in the semiclassical approximation, and
to show that (cf. \zaz) 
\eqn\vov{
A_2 (\xi) \ \sim\ \   |\xi_1 - \xi_2|^{2  \g -4}  \  , \ \ \ \ \ \ \ \
\g= \g (E, S)  \ ,  }
where $\g$ is approximatelt the same 
as in \maar, i.e.   $ \g \approx  - S  - \ha \a'   E^2$
so that demanding the  2-d conformal  invariance implies the leading Regge trajectory 
relation,  $ E = \sqrt { {2\ov  \a'} S}$.
The idea  is  again 
that in the  limit  when  both $E$ and  $S$  are of order of  the string tension
the path integral   representation for $A_2$ in \vvvv\
 should be dominated  by  a (complex)
classical trajectory with the  source terms (or boundary conditions)
determined  by the vertex operator insertions.

Note that the Regge trajectory relation  may be  rewritten as 
 $ \a' E = \sqrt { {2} \a' S}$  and thus  remains the same in the limit 
when  both $E$ and $S$  are 
of the order of string tension $ {1 \ov 2 \pi \a'}$. 
Indeed, as is well known, the  same relation  is obtained  for a 
classical string solution describing  folded closed string
 rotating about its center of mass. 
One finds  from the  closed string 
equations  defined on a 2-cylinder  ($\tau, \s\equiv \s + 2 \pi$)
with Minkowski signature in both target space and
world sheet
\eqn\flat{
t = \k \tau \ , \ \ \ \ \
 X= x_1 + i x_2 = \ r(\s) \ e^{ i \phi(\tau)} \ ,
\ \ \ \  \bX= x_1 - i x_2
= \ r(\s) \ e^{- i \phi(\tau)} \ ,  }
\eqn\fld{
\phi(\tau) = n \tau \ , \ \ \
\ \ \ \ r(\s) = \w   \sin (n \s) \ ,  \ \ \ \ \ \
\w= {\k\ov n} \ , \ \
 \ \ \ \ \  n=1,2,... \ . }
The relation between $\w$ and $\k$  follows from the
 conformal gauge constraint. 
The energy and the spin are 
\eqn\eess{
E= { 1 \ov 2 \pi \a'} \int^{2\pi}_0 d \s\ \dot t =
{ \k  \ov \a'} \ , \ \ \ \
\ \  S= { i \ov 4 \pi \a'} \int^{2\pi}_0 d \s\
( X \dot  \bX - \bX \dot X)= {n \w^2 \ov 2 \a'}  \ , }
so that for   the leading Regge trajectory state (having 
minimal energy for a given spin) one has
 $n=1$,  i.e. a single fold of the closed string. Then  
  $E= \sqrt {{ 2 \ov \a'} S}$. 
  \foot{Another classical solution describing 
  oscillating  closed string  is formally obtained from \flat,\fld\ by
  interchanging $\tau$ and $\sigma$ in the expression for $X$, 
  i.e. $\phi = n \sigma$, $r= \om \sin n \tau$. In this case 
  the spin is equal to zero, while we still get 
  $ E= \k/\a'$ and $ \om=\k/n$.  The vertex operator 
  representing such state is  a scalar 
  operator  from level $n$:  $e^{-iEt} \del^n  X {\bar \del}^n
   \bar X$. Similar solutions in AdS case were constructed in 
   \mina.}


The explicit form of  the correlator  \vvvv\  computed in conformal gauge 
on a  2-sphere or complex plane  is (we rotate to the Euclidean time
$t_e= i t $)\foot{In  our notation $\xi = \xi^1 + i \xi^2,\   \bar \xi = \xi^1 -
i \xi^2,\ d^2 \xi = d \xi^1 d \xi^2,$
and $ \del = \ha ( \del_1 - i \del_2) , \
\bar \del = \ha ( \del_1 +  i \del_2) .$
The Green's function of the Laplacian
$\del^a \del_a = 4 \del \bar \del$
on the complex plane is
$G(\xi-\xi')= - { 1 \ov 4 \pi} \ln |\xi - \xi'|^2 $,
i.e.
$- 4 \del \bar \del  G (\xi) = \delta(\xi)$.
Here $\delta (\xi)$ is the 2-d delta-function normalized  in such a way  that
$\int d^2 \xi \ \delta(\xi)=1$, so that
$ \bar \del { 1 \ov \xi } = \pi \delta (\xi)$,
etc.
}
$$ A_2 (\xi_1-\xi_2)
= \int [dt_e dXd\bar X... ]\
 \   e^{ - { 1 \ov  \pi \a'} \int d^2 \xi
\ ( \del t_e \bar \del t_e + \del X \bar \del \bar X  + ...) }  $$
 \eqn\ivv{
\times \   [ e^{- E t_e }
 \left( \pa {X } \bar{\pa} { X }
 \right)^{S/2} ] (\xi_1) \
[   e^{ E t_e  }
 \left( \pa {\bar X } \bar{\pa} { \bar X }
 \right)^{S/2}  ](\xi_2) \
.   }
This is a gaussian integral that of course can  be   easily
computed exactly, giving \vov\   with the coefficient  $\g$ in  \maar. 

In the case of  the ``classical''  values of 
 $E$ and $S$  we may instead apply 
the semiclassical method to compute $\g$.
We start by  rewriting  the  vertex operator
 insertions in \ivv\ in the exponential form as
\eqn\eex{
\exp \big( - E \big[t_e (\xi_1) - t_e (\xi_2)\big]\big) 
\ \exp \big[   \ha S \big( \ln \left[ \pa {X } \bar{\pa} { X }
  (\xi_1) \right]
+   \ln \left[ \pa {\bar X } \bar{\pa} { \bar X }
  (\xi_2) \right] \big) \big] \ , }
and then look for  a  stationary point of the total
expression in the exponent in \ivv,
 i.e. of the sum of the classical action  with  the terms in
the \eex, i.e.  $ O({1\ov \a'}) + O(E) + O(S)$.

While the contribution of the time coordinate is the
obvious one
(i.e. the same as in the case discussed in section 2.2),
the solution for $X, \bX$ may seem to be  less trivial to find, 
 given a  complicated non-linear form of the
terms coming from the vertex operator
insertions.  However, one may   expect 
that  details of these insertions should not be too 
important: the 
they  should just  select the relevant  solution of the free
Laplace equation for $X, \bX$ which effectively
 describes a classical trajectory associated with a
spin $S$ configuration.
If the  role of the ``sources''  is only 
 to specify appropriate boundary conditions
(though different from simply
$\delta$-functions   appearing in the case
of the external momentum insertions like $e^{-Et}$ one)
a natural guess is  that the associated stationary point
solution should be a complex analog
of the real Minkowski-time  rotating 
folded   string solution \flat\  with  spin $S$.

Using  translational invariance  we may set 
$\xi_1=0$.
Then the equations  obtained by varying $t$, $X$  and $\bX$ in
the effective action in \ivv,\eex\  are
\eqn\tat{
\del \bar \del t_e
= \ha \pi \a' E  [  \delta (\xi) -  \delta (\xi-\xi_2)] \ , }
\eqn\ua{
\del \bar \del \bX
= \ha \pi \a' S \big( \del [ {1 \ov \del X} \delta (\xi)]
+ \bar \del [ {1 \ov \bar \del X} \delta (\xi)] \big)
\ ,  }
\eqn\uas{
\del \bar \del X
=\ha \pi \a' S \big( \del [ {1 \ov \del \bX} \delta (\xi-\xi_2)]
+ \bar \del [ {1 \ov \bar \del \bX} \delta (\xi-\xi_2)] \big)
\ .  }
The relevant solution of this system  can be constructed  by 
 starting from
  the  known classical solution \flat,\fld\ on the 2-cylinder.\foot{The
  argument that follows was worked out in collaboration with S. Frolov.
  A related derivation of the marginality condition
  using  a single vertex operator insertion and 
  imposing the conformal gauge constraint 
   was given   K. Zarembo.}
We shall first
 note that the conformal transformation of the complex plane 
$ \xi \to \zz$ defined by 
\eqn\zaze{
{ \zz^{-1}} = {  \xi^{-1} }  - { \xi_2^{-1} }  }
maps the points $\xi=\xi_1=0$ and $\xi=\xi_2$ to points $\zz=0$  and
$\zz=\infty$, respectively. Then the complex $\zz$-plane with punctures 
at $0$ and $\infty$ can be mapped 
(as in the context of the standard  operator -- state correspondence) 
to a Euclidean cylinder $(\tau_e,\s)$  by 
the transformation \euc, i.e. 
\eqn\taaak{
\zz = e^{\tau_e + i \sigma}\ , \ \ \ \ \ \ \ \   \tau_e = i \tau  \ . } 
The classical solution \flat\
on the Minkowski 2-cylinder  can be 
rotated  to the Euclidean solution  ($t \to -i t_e, \ \tau \to
-i \tau_e$) which takes the following form in terms of  $\zz$:
\eqn\zaq{
t_e = \ha \k \ln |\zz|^2 \ , \ \ \ \ \ \
X= {\w \ov 2 i } ( \zz - \bar \zz)  \ , \ \ \ \ \
\bX ={\w \ov 2 i } ( {\bar \zz}^{-1} -  \zz^{-1})   \ . }
We have  introduced a parameter
$\w$ which is equal to $\k$ on
the classical single-fold solution \fld\ satisfying the
 conformal constraint.\foot{
Note that  $X$ and $\bX$ in the Euclidean solution  \zaq\ 
 are no longer related by a complex conjugation.
Instead, they  are related by the transformation $\zz \to { 1 \ov
\bar \zz}$ or $\tau_e \to - \tau_e$, which, as expected,  interchanges the two
insertion points, or reverses  the time axis on the cylinder.}

Transformed to  the $\zz$-plane the system \tat,\ua,\uas\ has the same form 
with $\xi\to \zz$ and $\xi_2  \to \infty$.
  Then $t_e$, $X$ and $\bX$  in \zaq\ solve the free
Laplace equation  on
the $\zz$-plane apart from the points $\zz=0$ and $\zz=\infty$.
 At these special points they are indeed
supported by the  source terms in the right hand side of \tat,\ua,\uas\ 
provided  $\k$  and $\w$ in \zaq\  are  related to the  ``source'' 
coefficients $\a'E$ and $\a'S$  in  \tat,\ua,\uas\  by 
\eqn\sdf{ \a' E= \k \  , \ \ \ \ \ \ \     \w^2 = 2 \a' S     \ . }
These are the same  relations as the ones  \eess\  for the 
classical solution \fld, but here obtained in a very  different way. 

The solution of the original system \tat,\ua,\uas\
is then given by \zaq,\sdf\   with $\zz$ replaced by $\xi$ according to
 \zaze.
The final step is  to evaluate the effective  action in the path integral \ivv,\eex\ 
on this Euclidean solution,  i.e. find the coefficient 
$\g(E,S)$   in  \vov\  and thus  determine $E= E(S)$ 
from the conformal invariance condition  $\g=0$.
Following the discussion in  section  2, the time coordinate  contribution 
 is given, as in the standard gaussian integral,  by the 
1/2 of the external point  contribution. 
One finds 
that the $(X,\bX)$  part of the classical string action
vanishes on the solution 
so
that  the  semiclassical contribution
 to the $(X,\bX)$ path integral comes only from the $S$-dependent 
vertex operator insertion terms in \eex\ evaluated on the solution.
The latter is essentially
 determined  by the
{\it canonical} dimension of the vertex operators.\foot{As always 
for  string excited states in flat space, the mass shell 
condition expresses the 
 balance between  the anomalous dimension of the $e^{ipx}$
operator and canonical dimension of the pre-exponential
$\del X$ factors. The spin $S$ term in \maar\ represents 
   the canonical dimension.}
Differentiating $X$,$\bar X$  in \ivv\ or \eex\
over $\xi$ in \zaze,
 we get  (regularising and omiting some  obvious singular terms)
the factor $ |\xi_2|^{-2 S} $, which,
combined with the $t$-contribution,  then  reproduces
\vov\ with $\g= \ha \a' E^2 -  S$, i.e. 
gives  the leading Regge trajectory  relation.

To summarize:

(i)  The role of matching onto the  source terms
 in the classical
equations \tat--\uas\ is to relate the parameters $\k,w$ of the semiclassical
trajectory  to the quantum numbers $E,S$ of the vertex operators.
In the
classical rotating string solution \flat\ this   was done 
 by
fixing the values \eess\ of the conserved quantities on the solution.

(ii) 
The  role of the  marginality condition $\g(E,S)=0$
is to relate $E$ to $S$. In the classical solution 
this  followed  from the  conformal
 constraint that related  the parameter $\k$  determining  the energy $E$
and  the parameter $\w$  determining  the spin $S$.

\newsec{Vertex operators for spinning string states in  $AdS_5$
 and $S^5$ }

One would like to try to  give a similar semiclassical 
path integral derivation 
of dimensions of  string states  with large \cla\ 
 spin $S$ in $AdS_5$ or intrinsic 
angular momentum $J$ 
in $S^5$ with 
  for which the classical 
solutions of \gkp\  predict that 
\eqn\pred{
\D(S)_{{S\ov \sql} \gg 1} = S + {1 \ov \pi}  \sql \ln S + ... \ , \ \ \ \ \ \ 
  \D(J)_{{J\ov \sql} \gg 1} = J + { 2 \ov \pi}  \sql + ...    \ . } 
The first step  should be to identify the corresponding 
vertex operators  
that should be analogs  of the operator 
with orbital momentum \tyu.
One could then attempt to 
 compute the 2-point functions like \vvvv,\vov \  
in semiclassical approximation   to determine  the anomalous power 
$\g=\g(\D,S)$ or $\g=\g(\D,J)$,  
 which, when set to zero, should be reproducing hopefully 
  the results \pred. 
  
  Here we  will not be able to complete the second part of this program
  (though we will make some comments on it in section 4.2), 
  and  will most  concentrate (in sections 4.1 and 5)
   on the first part, i.e.
   determining the 
  structure of vertex operators,    which 
  already turns out to be quite non-trivial. 
  
\subsec{\bf General structure of vertex operators}

The natural  conditions on  such  vertex  operators 
are: (i) they should carry the required  quantum numbers $\D$ and $S$ or $J$,
i.e. should have appropriate transformation properties under $SO(1,5)$ 
and $SO(6)$;
(ii) in the flat-space limit they should reduce to the 
leading Regge trajectory  vertex  operator  \fle. The latter condition 
implies that their level number or  canonical 2-d  dimension 
(equal to the number of 2-d derivatives)  should be 
$S$ (or $J$).\foot{As above, we shall be 
 ignoring 
terms depending on the fermionic string coordinates 
(as well as possible $\a'$-corrections). A possible excuse is 
that one  expects that  they should not be important in  
the leading semiclassical 
approximation (fermions were  of course  set to zero 
in the classical solutions of \gkp).}

An  obvious candidate  for a  vertex operator representing a string state
with angular momentum $J$ in $S^5$  is then (cf. \tyu,\fle)\foot{In 
flat space limit $n_6\to 1, \ n_1 \to  \theta 
\cos \vp, \ n_2 \to \theta \sin \vp $, etc., with $\theta$ playing the
role of a radial coordinate.}
\eqn\jjj{
\V_J(\xi) = c (N_+)^{-\D}\big( \pa n_x \bar{\pa} n_x \big)^{J/2}\ , 
\ \ \ \ \ \ \ \ \     n_x \equiv  n_1 + i n_2  \ .  }
The  analogous  operator   for a spinning string  in $AdS_5$  is 
\eqn\sss{
\V_S(\xi) = c (N_+)^{-\D}\big( \pa N_x \bar{\pa} N_x \big)^{S/2}\ , 
\ \ \ \ \ \ \ \ \     N_x \equiv  N_1 + i N_2  \ .  }
It may be useful to record also the 
explicit form of $\V_S$ in terms of independent  coordinates  on $AdS_5$.
The Euclidean  $AdS_5$ sigma model Lagrangian 
$${L} = -\pa N_+\bar{\pa}N_- + \pa N_\m \bar{\pa}N_\m \ , \ \ \ \ \ \ 
N_+ N_- - N_\m N_\m =1  \ 
$$ can be written as 
\eqn\lasso{
 L=  \del T \bar \del T
+ \pa N_\m\bar{\pa}N_\m
+
( \del \bar \del T  + \del T \bar \del T) N_\m N_\m   \ , \ \ \ \
 \ \ \    \ T \equiv \ln N_+ \ ,  }
or as 
$$
{ L} = (1+N_\m N_\m ) \pa \tt \bar{\pa} \tt
- {\pa (N_\m N_\m )\bar{\pa}
(N_\n N_\n )\ov
4(1+ N_\l N_\l )} +\pa N_\m\bar{\pa}N_\m\  $$
\eqn\asd{
= \   (1+N_x \bar N_x ) \pa \tt \bar{\pa} \tt- {\pa (N_x \bx)\bar{\pa}
(N_x \bx)\ov
4(1+N_x \bx)} +\pa N_x \bar{\pa} \bx  + ...\ ,    }
where   dots stand for the  terms depending on
the remaining coordinates
 $N_3, N_4$.
Here $\tt$ 
is  the  global ``Euclidean  AdS time''
variable\foot{Note that this $\tt$  is not the same as 
Euclidean rotation of 
$t$ used in  a footnote below \nen: the  Euclidean
global coordinates
 here are  related to the direct Euclidean analog of the coordinates 
there by a global $O(1,5)$ transformation which   
effectively interchanges
$N_0$ and $N_4$ (so that now $N_4= \cosh \rho \ \sin t$,
$N_0= \hat n_4 \sinh \rho$).
 The relation to the Poincare coordinates is  however  is
the same.} 
\eqn\tim{
\tt  \equiv  \ln N_+ - {1\ov 2}\ln (1+N_\m N_\m )=
T- {1\ov 2}\ln (1+ N_\m N_\m ) \ , \ \ \ \ \ \ \ \
e^{2\tt  } = { N_+ \ov N_-}    \  . }
In these coordinates  \sss\   becomes 
\eqn\vevo{
\V_S (\xi)  =\  c\ e^{-  \Delta T } ( \del N_x \bar \del N_x)^{S/2} 
= \ c\ e^{ - \Delta \tt} (1 + N_\m N_\m )^{-\D/2}  
 ( \del N_x \bar \del N_x)^{S/2}  \  . } 
Written  in terms of the Poincare coordinates 
this vertex operator takes the form (cf. \fle) 
\eqn\vevo{
\V_S(\xi)  = c \ \big( { z^2 + x_\m x_\m \ov z}\big)^{-\D}
\big[ \pa ({X\ov z}) \bar{\pa} ({  X \ov z})
 \big]^{S/2}\ ,  \ \ \ \ \ \ \ X= x_1 + i x_2   \ . 
 }
While  the vertex operators \jjj\ and \sss\  appear to be natural choices, 
there remains a  question then if  they are indeed the correct ones 
at the quantum  \adss string level, in particular, whether  they  
mix with other similar operators.  That would mean that they   
represent  eigen-operators of the anomalous dimension 
matrix, or,   equivalently,  that the associated 
wave functions solve the corresponding  ($\a'$-corrected)
 KG equation.

It turns out that, in contrast to what was in the case of 
the spinning state operator \fle\ in flat space, 
 the operators \jjj\ and \sss\ 
do mix with other similar operators   already in the 1-loop 
$O({1 \ov \sql}$) approximation for the \adss sigma model.
At the same time, the operator \tyu\ representing a 
point-like string state with an  orbital momentum $J$ 
is indeed an eigen-operator of the anomalous dimension matrix. 

As we shall explain  in  section 5, for any value of $J$ 
the operator $(\del n_x \bd  n_x)^{J/2}$ of 
(bosonic) sigma model on a sphere $S^{\N-1}$  mixes under 
the 1-loop renormalization 
with the following operators 
 (all of which have the same canonical dimension 
equal to the angular momentum  $J$)
\eqn\mii{  (n_x)^{2p + 2q  } (\del n_x)^{J/2 - 2p}   (\bar
\del n_x)^{J/2 - 2q }
( \del n_m \del n_m)^p ( \bd n_k \del n_k)^q  \ , \ \ \ \ \
p,q= 0,..., J/4  \ , } 
where there are sums  over the indices $m$ and $k$
with range $1,...,\N$ (in the 5-d  case of our interest 
$\N=6$).  
A similar but somewhat more complicated mixing  takes place 
for the spin $S$ operator \sss\ in the bosonic $AdS_{\N-1}$ sigma model --
the operator \sss\  may mix with 
\eqn\enek{
 N_+^{- \D - p-q} N_x^{p+q} (\d N_+)^p (\d N_x)^{S/2-p} 
(\bd N_+)^q (\bd N_x)^{S/2-q} + ...  \  ,  
\ \ \ \ \      p,q= 0,..., S/4  \ , }
where  dots  stand for terms 
involving powers of  scalar operators
 $ \del N_M \del N_M$ and  $\bd N_K \bd N_K$
similar to the ones in \mii. 
Again,  these operators have the same $SO(1,5)$ quantum 
numbers $(\D,S)$ as \sss.\foot{Note  that we do not include 
extra real factors 
like $(\del n_m \bd n_m)^r$ or  $(\del N_M \bd N_M)^r$.
Their maximal power $r$ plays the role of an additional 
quantum number, and their (1-loop) 
renormalization  always decreases $r$.}

This obviously raises the question of whether this mixing is
important  to take into account in an attempt of a semiclassical 
derivation of the corresponding anomalous dimensions  for large values 
of quantum numbers.
The  linear combinations  of the above operators 
that represent the true spinning string  vertex operators 
 may simplify in the limit of large quantum numbers.
We are unable to answer this question
here,  and in the next subsection  will  only 
sketch a  strategy of   possible
semiclassical argument in the $S^5$ rotation case 
starting simply with the ``naive'' form of the corresponding 
vertex  operator \jjj.
The $S^5$ rotation case is more transparent than the 
spinning $AdS_5$ case
but is already quite non-trivial to illustrate the main 
points.

\subsec{\bf Towards a semiclassical  derivation 
of anomalous dimension of 
vertex operator with  $S^5$ angular momentum }

In the semiclassical approximation 
the $AdS_5$ and $S^5$ factors $N_+^{_\D}$ and $(\d n_x \bd
n_x)^{J/2}$ in \jjj\  can be treated independently.
As in the  $S^5$ orbital momentum case of section 2.2 
the $AdS_5$ factor contributes to the 2-point correlator 
 the same term  as in \zaz\ 
with $\g(AdS_5)= { 1\ov 2\sql} \D^2$. 
For $\xi_1=0$ the stationary point 
is as in \zaq, i.e. 
\eqn\zaq{
t_e = { 1\ov 2} \k \ln |\zz|^2   \ , \ \ \ \ \ \ \   \k = {\D \ov
\sql} \ , } 
where $\zz$ is given by \zaze. 
 The main complication lies in the $S^5$ sector. 
 Since the vertex operator depends only on $n_x$ components of 
 6-vector, we may assume that  $n_4=n_5=n_6=0$   and consider the 
 $S^2$ subspace of $S^5$  parametrized by 
 $n_x, \bnx$ and $n_3= \sqrt{ 1 - n_x \bnx} $. 
 Then the relevant part of the  \adss string   action 
 will  be the $S^2$ sigma model  (cf. \asd) 
 \eqn\accd{ 
 I  = { \sql \ov \pi} \int d^2 \xi\ 
 [  \d n_x \bd \bnx  + { \d ( n_x \bnx)  \bd (n_x\bnx)
  \ov 4 ( 1- n_x \bnx) }] \ . } 
Adding to this action the (minus) logarithm of 
the product of the two  operators $\V_J$ and  $\V_{-J} =
V_J(n_x \to \bnx)$ at points $\xi_1=0$ and $\xi_2$, 
i.e. 
 \eqn\sosk{- \ha J \ln [ (\d n_x \bd n_x)(0) (\d \bnx \bd \bnx)(\xi_2 
 )]   } 
 and assuming 
 that the angular momentum $J$ is of order of 
 the string tension $ \sql$, 
 we can then try to find   a (complex) 
 semiclassical trajectory $n_x(\xi)$ 
 that  saturates this path integral in the semiclassical 
 approximation. This would then 
 determine the $S^5$-factor contribution  
  to $\g$ in the analog of \zaz.

 In view of the discussion of the flat space case in section 3
 it is natural to anticipate that the stationary point solution will
 be
 related by an analytic continuation and 
 the ``cylinder $\to$ complex plane''  map 
 to the classical solution in \gkp\ describing folded closed string
 rotating  at the north pole of $S^5$. 
 In Minkowski notation this solution has 
 the following form on the 2-d
 cylinder (here the metric of $S^5$  taken to be 
$ds^2 = d \theta^2 + \sin^2 \theta\  d \vp^2 + \cos^2 \theta\  d
\Omega_3^2$)
\eqn\sool{
  \vp = \om \tau,\ \ \ \ \ \  \theta=\theta(\s), \ \ \ \ \ 
\theta'' +  \om^2 \sin \theta\ \cos \theta =0  \ . }
We should not impose the conformal  gauge constraint 
that in \gkp\ related $\om$ to $\kappa$ in $t=\kappa \tau$ 
and thus $J$ to $\D$:
here this relation should 
 follow from the marginality condition of the
vertex operators, 
i.e. from the vanishing of the coefficient $\gamma$ in the 2-point
correlator \zaz.
Using the parametrization
\eqn\parrm{ 
n_x = \sin \theta \ e^{i\vp} \ , \ \ \ \ \ \ \ \ \ \ \  
\bnx = \sin \theta \ e^{-i\vp} \ , }
rotating the solution to the  euclidean one, $\tau \to - i \tau_e$, 
and mapping the cylinder onto the complex plane according to 
\zaze,\taaak\   we end up with the following complex solution 
for the independent functions $n_x$ and $\bnx$ on the
$\zz$-plane:\foot{Note
that $n_x \to \bnx$ under $\zz \to \bar \zz^{-1}$ 
in agreement with the  exchange symmetry of the two vertex operators
or time reversal on the cylinder.}
\eqn\exch{
n_x = \ f({\zz \ov \bar \zz})\ |\zz|^{\om} \ , \ \ \ \ \ \ \ \ \   
\bnx = \ f({\zz \ov \bar \zz})\ |\zz|^{-\om} \ ,  } 
where $f$ satisfies the equation (the transformed form 
of   the equation for $\theta$ in \sool)  
\eqn\soop{
(1-f^2) f''  + f f'^2  +  \om^2 f (1-f^2)^2 =0 \ , }
$$
f' = { d f \ov d \s} \ , \ \ \ \ \ \ \ \ \ 
 \s= - { i \ov 2} \ln { \zz \ov \bar \zz}
\ . $$
Eqs. \exch,\soop\ represent 
a particular   vacuum solution of the $S^2$ sigma model \accd.\foot{
 The solution  in the case of the 
spinning string state in $AdS_5$ (i.e. the one that should be relevant
for the semiclassical computation of the 2-point function of the
operators \sss) has a similar structure:
$N_x = \ F({\zz \ov \bar \zz})\ |\zz|^{\om} $, \    
$\bar N_x  = \ F({\zz \ov \bar \zz})\ |\zz|^{-\om} $, 
where  $F= \sinh \rho (\s)$, with 
$ \rho''= ( \k^2 -\om^2) \sinh \rho \cosh \rho$.}
The  next and main  
 issue is then to show that matching it onto the source 
terms  coming from \sosk\ 
we get the required relation between the angular momentum $J$
and 
the parameter $\om$ of the solution.\foot{For this it is sufficient 
to consider just a single vertex operator insertion.} 
 The explicit  form of this  relation 
is in general quite complicated \gkp\  
but it should at least simplify  in the limit of large $J$.
 Unfortunately, the non-linear form of the equation \soop\
 makes matching onto $\delta$-function siurce terms 
 (cf. \ua,\uas)  hard to  implement explicitly.
To relate $J$ to $\om$ in the limit ${J\ov \sql }\gg 1$ 
 it  may be sufficient to  use only  an 
asymptotic form of the solution \exch, with introduction of 
a 2-d cutoff near $\xi=\xi_1$ or $\xi_2$. 

As in section 3, the final step will be to evaluate the sigma model
action (plus source terms) on the solution \exch\
to find the $J$-dependent contribution to the coefficient $\g$ in 
the analog of \zaz, and then hopefully reproduce 
$\D = J + ...$ in \pred\ from the conformal invariance 
 requirement that $\g=0$. 

One may object the utility of this method, given that  it seems 
to be  much more complicated and  indirect as 
  compared to the original  
 (``semiclassical string states on a cylinder'') 
 approach of \gkp.\foot{In particular, it seems unlikely that one 
 will be able to derive  the full relation $\D(J)$ in \gkp\
 for any value of $J$ in the present vertex operator 
 approach.}
 Still, we hope that 
 further understanding of this method may shed more light on
 details of the  ``string state -- vertex operator''  correspondence 
 in the nonlinear sigma model cases  like the present one.

\newsec{Renormalization of composite operators  in 
$S^{\N-1}$ and $AdS_{\N-1}$  sigma model }

Let us now return to the issue of the mixing 
of the rotation vertex operators, i.e.  \jjj\ mixing 
 with \mii\ and  \sss\ with \enek.

Having  in mind applications to  the 
 $AdS_5 \times S^5$ string theory,  one may expect that in the  1-loop 
approximation one may ignore the fermions  and thus, in particular, 
treat  renormalizations in the  $AdS_5$ and $ S^5$ sectors  
as independent. However, as already mentioned above, 
this is not  a priori obvious. Indeed, 
the superstring action \mets\ contains the R-R 5-form coupling term  
which has    structure  $ \theta \theta ( \del n + \del N) $.
String vertex operators  may also contain fermionic terms
like $ ( \del n + \theta \theta)(...)$  and $
 ( \del N + \theta \theta)(...)$.  Pairing the fermionic terms in the action
 and in the vertex operators one may thus get 
 additional fermionic contributions to 
 the renormalization of the vertex operators. 
One   may also get, via such fermionic contributions, extra  mixings between 
the operator factors  from the $AdS_5$ and from the $ S^5$.
Here we shall not attempt to take 
such fermionic contributions into the account, 
 hoping  that 
they would  not qualitatively change our main conclusions.

Extending  the discussion in \pol, in this section  we shall study 
 mixing and anomalous dimensions  for  vertex-type  
 operators in the $O(\N)$ invariant  2-d sigma model, 
applying  the   general method   described in  \weg\ (see also \cast). 
As in  \refs{\weg}, we  will  
 consider explicitly only the 1-loop approximation in bosonic 
sigma model, but  we expect  that  these  1-loop results may 
 clarify  some important  features  of the general  case. 
The  discussion  of the $S^{\N-1}$  sigma model case can be 
generalized   also to  the $AdS_{\N-1}$ case 
using a simple continuation rule as in \pol.


Let us start with some general remarks. 
Given a sigma model\foot{In general, the sigma model 
should be defined on a curved 2-d background; one 
should include all possible 
couplings to the curvature of the 2-d metric and 
instead of usual scale invariance 
demand the vanishing of the Weyl anomaly 
coefficients. 
In addition to Klein-Gordon-type equation for the tensor 
function,  that leads, as in flat space \wein, 
 also to divergence-free and tracelessness
conditions  (see, e.g., \buch\ and refs. there). }
with the action $S = { 1 \ov \pi \a' } \int d^2 \xi\  G_{mn}(x) 
\del x^m \bd x^n $  one may perturb it by an arbitrary 
higher-dimensional local operator of  the type 
(with total number of 2-d derivatives or level number $Q$)
$V(f) = f_{m_1 ...m_j} (x) \del^{k_1} x^{m_1} .... {\bd}^{k_h} x^{m_j} $
and compute the renormalization of the coupling  tensor 
$f_{m_1 ...m_j}$. Demanding the vanishing of the 
corresponding 
beta-function to linear  order in $f_{m_1...m_j}$
(but to all orders in $\a'G^{mn}$) 
determines the corresponding 
 marginal string vertex operator \calg. 
 The resulting equation  has symbolic form 
 $ \hat \gamma  f = [2- Q +  \ha \a'  \nabla^2 + \a' R_{....}  + 
  \sum \a'^k R....R \nabla...\nabla ] f =0 $,
 where $\hat \gamma$ is the ``anomalous dimension'' operator
 (in which we included also the canonical dimension term).
 For example, a  2-nd rank tensor $f$ (massive spin 2) 
 satisfies a Lichnerowicz-type  equation \buch.
  Solving such  equations  for $f$  is equivalent to finding the
  the eigenvalues and eigen-vectors 
  of the anomalous dimension operator. 
  Unfortunately, the general form of $\hat \g$ operator 
  for generic tensor $f$ and metric $G$  is not known  even 
  to the leading (1-loop) order in $\a'$.
  For this reason,  in most cases
  (with possible exceptions being WZW models or 
  some plane-wave type models) 
   one is not able to  use a universal expression for $\hat \g$ 
   and needs to  calculate 
  the anomalous dimensions of specific higher-derivative 
    operators  starting from  ``first principles''. 

  When the sigma model metric $G_{mn}$  has a  large
   global symmetry
  (e.g., corresponds to a symmetric space)
  it is natural to construct the operator  $V(f)$ so that it is 
  covariant under the  global symmetry. In particular, instead 
  of using  local coordinates $x^m$,  we may  write $V(f)$ in terms 
   some   global coordinates on which the group acts linearly. 
  This is what we will do   in the discussion below.

\subsec{\bf Rules of 1-loop renormalization in $S^{\N-1}$   sigma model }

To set up the notation, let us   start with the action of the    $S^{\N-1}$ \sm  action
defined on a complex plane
\eqn\siga{
S = { R^2 \ov 4\pi \a' } \int d^2 \xi\ ( \del_a n_m)^2 
=  { \sql  \ov  \pi  } \int d^2 \xi\  \del n_m \bd n_m   \ , \ \ \ \ \ 
n_m n_m =1  \ , } 
where  $m=1,..., \N$ and  $ \ge \equiv 
{1 \ov  \sql } = { \a' \ov R^2} $ 
plays the role of the standard  dimensionless \sm coupling constant, 
which runs according to the   well-known RG equation\foot{The full 
 \adss string \sm  is of course conformally invariant 
due to additional  fermionic contributions 
to beta-functions of each of the $AdS_5$ and $S^5$ coupling constants 
\mets\ 
coming from the R-R 5-form coupling term.
 } 
\eqn\runn{
\dot \ge   = - \epsilon  \ge 
+ (\N-2)  \ge^2  +  (\N-2)  \ge^3  + ... \ , \ \ \ \ \ 
\ \ \ \ \    \ge \equiv   { 1 \ov \sql} \ , \  \ \ \ 
    \epsilon = d-2  \ .   }
In view of the   discussion in the previous section 
 we shall be interested in renormalization of the 
following class of $O(\N)$ covariant 
operators\foot{Since eventually 
we will be interested in operators corresponding to 
states on leading  Regge trajectory we 
do not include terms with multiple derivatives.} 
 \eqn\jij{
O_{\el,s} =  f_{k_1...k_{\el} m_1...m_{2s}}  
n_{k_1} ...  n_{k_\el}  \d n_{m_1} \bd n_{m_2} ...
\d n_{m_{2s-1}} \bd n_{m_{2s}}
 \ . } 
These form  a special  class of composite 
operators considered in \weg\ (see sects.  8,9 there). 
We shall  assume that $\N$ is even and thus we can split  
the components of $n_k$ into complex  pairs, i.e.
\eqn\jkl{
n_x = n_1 + i n_2 \ , \ \ \  \bnx = n_1 - i n_2 \ , \ \   \ \ \ \ 
n_y = n_3 + i n_4 \ , \ \ ... .  } 
In the  case of  Euclidean $AdS_{\N-1}$  \sm 
 the corresponding  split 
of  the $SO(1,\N-1)$  unit vector 
$N_M$  is into $ N_+,N_-,N_x,...$, where $N_\pm$  are real, 
i.e. $N_+N_- - N_x N_{\bar x} - ... =1$. 

To compute the  1-loop renormalisation of the operators \jij\ one may 
follow the general method of \weg\  and use 
 the following results for ``pairing'' of 
different building blocks of 
composite  operators\foot{We keep only the singular term 
proportional to $I$ and use the notation of 
\weg: $ < A B> $ $ = <A> B + A <B>  + <A,B>$,  where 
 $A$  and $B$ are some operators; if they depend on $n_k(x)$ only 
then $ <A,B> = \int d^2 \xi  d^2 \xi'  <n_k (\xi), n_m (\xi')> 
 { \delta A \ov \delta  n_k (\xi)}   { \delta B \ov  \delta n_m (\xi')} $, etc.
Also,  $ < A(n)> = \ha  \int d^2 \xi d^2 \xi'$ $  <n_k(\xi) , n_m(\xi')> 
 { \delta^2 A \ov \delta  n_k (\xi)  \delta n_m (\xi')  } $.
These pairing rules can be understood also in the 
framework of the   background field method \cast.
Note  that in the process of renormalization 
one is   allowed  to ignore redundant operators proportional to 
equations of motion, e.g., 
$
[ \d \bd n_k - n_k ( \d n_m \bd n_m) ]  C_k (n)$, 
where  $C_k$ is any local operator. } 
\eqn\asd{
<n_k >  = - \ha (\N-1) I n_k \ , \ \ \ 
< n_k , n_l> = - I( n_k n_l  - \delta_{kl}) \ , 
\ \ \  \ \ \ \   I= - { 1 \ov 2 \pi \epsilon}  \to  \infty\ , 
}
 \eqn\opp{
<n_k, \d n_l> = - I \d n_k n_l \ ,\  \ 
 <n_k, \bd n_l> = - I \bd n_k n_l \ ,\  \  }
   \eqn\boom{
 <\d n_k, \d n_l> =  I n_k n_l \d n_m \d n_m \ , \ \ \ \ \ \ \ 
 <\bd n_k, \bd n_l> =  I n_k n_l \bd n_m \bd n_m \ , \ }
\eqn\dsa{  <\d n_k ,\bd n_l> = - I (\bd n_k \d n_l -\delta_{kl}  \d n_m \bd n_m )
\ . } 
Contracting indices $k,l$ in \dsa\    we get
$<\d n_k ,\bd n_k> =  I (\N-1) \d n_k \bd n_k $. Then 
 using $<n_k >  = - \ha (\N-1) I n_k $ we conclude that 
\eqn\nadd{   <(\d n_k \bd n_k)> = \d <n_k> \bd n_k  + \d n_k \bd <n_k> +  
<\d n_k ,\bd n_k>  =  0   \  . }
For other scalar contractions one finds \weg\ 
\eqn\coop{
 <(\d n_k \d n_k)> = - (\N-2) I  \d n_k \d n_k\ , \ \  \ \ \ \ \ \ 
 <(\bd n_k \bd n_k)> =  - (\N-2) I  \bd n_k \bd n_k\ , \ \ }
\eqn\qda{
 <(\d n_k \bd n_k) ,  (\d n_m \bd n_m) > = 
2 I [ (\d n_k \bd n_k) (\d n_m \bd n_m) - 
(\d n_k \d n_k) (\bd n_m \bd n_m)] \ , }  
\eqn\acs{  <(\d n_k \d n_k) ,  (\d_a n_m \d_b n_m) > = 0 \ , \ \ \ 
 <(\bd n_k \bd n_k) ,  (\d_a n_m \d_b n_m) > = 0 \ ,  } 
\eqn\mop{
< n_l ,  (\d_a n_k \d_b  n_k)> =0  \ , \ \ \ \ 
<\d  n_l ,  (\d n_k \d  n_k)> =0  \ , \ \ \ \
<\d  n_l ,  (\bd n_k \bd  n_k)> =0  \ ,   }
\eqn\onn{
<\d  n_l ,  (\d n_k \bd  n_k)> =
I [ \d  n_l   (\d n_k \bd  n_k) - \bd n_l  (\d n_k \d  n_k) ] 
 \ .   }
Here $\del_a= (\del, \bar \del)$,   and similar relations 
are valid for complex conjugate combinations. 
Some of the above relations simplify 
when written in terms of the complex combinations $n_x$ in \jkl\  (note that $\delta_{xx}=0$)
\eqn\xex{
< n_x , n_x> = - I n_x n_x \ , \ \ \ \ \ \ 
 <\d n_x ,\bd n_x> = - I \d n_x \bd n_x   \  ,  \ \ \ \   
 <(\d n_x \bd n_x)> = -  \N I  \d n_x \bd n_x   \  ,
}
$$  <\d n_x ,\d n_x> = I n_x n_x  \d n_m \d n_m   \  ,  \ \ \ \ 
 <\bd n_x ,\bd n_x> = I n_x n_x  \bd n_m \bd n_m   \  . $$

\subsec{\bf Anomalous dimensions of some  simple operators }

We are now ready to discuss some  examples of 
operators \jij. 
 The simplest ``lowest level''  scalar  operator 
$ f_{k_1 ...k_\el} n_{k_1} ... n_{k_\el}$  with 
traceless $ f_{k_1 ...k_\el}$ is transformed under renormalization into
itself (see \asd).  It has the same anomalous dimension  $\g$ as its 
simplest (highest-weight) 
representative:\foot{If $f_\el$ is the coupling 
constant that multiplies a composite  operator when it is added to the \sm 
action, the anomalous dimension is defined by 
$f^{-1}_\el \dot f_\el = \g_\el (\bt) = 2 + O(\bt)$.} 
\eqn\opim{
A_\el = (n_x)^\el  \ \ :
\ \ \   \ \ 
 \g=2  - \ha [ \el (\N-1)  +  \el (\el-1) ] \bt + O(\bt^2) 
= - \ha  \   \el(\el + \N-2) \bt   + O(\bt^2)  \ . 
} 
The two $O(\bt)$ terms came from the two terms in \asd\ 
(note that for $n_x^\el$ the number of 
2-point pairings is $ \ha l (l-1)$). 
The total  coefficient  
 $  \el (\el  + \N-2)$ is of course 
the value of the corresponding quadratic
 Casimir (this is a particular spherical harmonic, i.e.
 it solves the  Laplace 
equation on the sphere). 

As follows from \coop, the Lagrangian
operator $L=\d n_m \bd n_m$  has zero dimension.
 Also, eq. \mop\ implies that it does 
not mix with  powers of $n_k$, i.e. 
the dimension of $n_x^\el \d n_m \bd n_m$ is given simply 
by the sum of the corresponding dimensions, 
$\g  = - \ha    \el(\el + \N-2) \bt  + O(\bt^2)$.

Similar results for operators of $AdS_{\N-1}$ \sm are found 
by replacing $n^\el_x$ and $\d n_m \bd n_m $ by 
 $N^\el_+$ and $\d N_M \bd N_M $, taking  $\el = - \D$ and 
changing  the sign  of the coupling $ \ge $, i.e. of the \sm 
coupling.\foot{Note that the term 
$\d N_M \bd N_M $ (with signature $+-----$) enters 
the action with the opposite sign 
compared to the term  $\d n_m \bd n_m$; this   reflects
 the opposite signs of the curvatures of $S^{\N-1}$ and $AdS_{\N-1}$.}
That implies that the dimension of 
 $N^{-\D}_+ \d N_M \bd N_M $  is 
$\g  =  \ha    \D(\D - \N+2)  \bt \ + O(\bt^2)$.
Applying these remarks to the case of the ``dilaton'' operator 
\tyu\ we verify the expression \margi\ for its perturbative anomalous 
dimension ($\el=J$ for the sphere factor and $\N=6$). 

Note that according to \dsa\ 
$ \d n_m  \bd n_k < \d n_m, \bd n_k >   =0 $.
Thus the left-right  factorized 
operators $ ( \d n_m \d n_m)^p (\bd n_k \bd n_k)^q$ 
 do not mix with  operators containing $\d n_k \bd n_k$ factors.
In particular, 
the number of $\d n_k \bd  n_k$ factors never increases, so 
it can be used  as a quantum number 
to characterise the leading term in the corresponding 
 operator \weg. An example of such 
an operator  considered  in 
\rf{\kra,\weg,\cast} is 
\eqn\zzz{ (\d n_k \bd n_k)^r+ ... \ \ : \ \  \  \   \g= 2-2r  +    r(r-1) \bt 
+ 
[{ 2 \ov 3} r (r-1) (r- {7 \ov 2})  + r (\N-2) ] \bt^2 +
O(\bt^3) \ . }   
Here  the 2-loop correction was computed in \cast.
Dots following the operator  $(\d n_k \bd n_k)^r$
stand for other operators it mixes with, and $\g$ is the 
corresponding diagonal element of the anomalous dimension matrix.\foot{ 
Here (as in \opim)  one is to account  for 
the number of 2-point pairings  $ \ha r (r-1)$, but  
 there is extra -2 in \qda\ as compared to \asd. The dimension indeed vanishes for $r=1$ in agreement with the above discussion of the dilaton operator.} 
In more detail, the above pairing rules imply that a 
renormalization-invariant  linear combination of the operators 
of the same canonical dimension $2-2r$ 
 is given by \weg\  (we assume that $r$ is even integer)\foot{When $r$ is odd one is to consider 
$  c_0 (\d n_k \bd n_k)^{r} + ... + 
c_{[r/2]}
(\d n_k \bd n_k)   (\d n_m \d n_m \bd n_l \bd n_l)^{[r/2]}$.}
\eqn\taps{  O_r= \sum_{p=0}^{r/2} c_p  
 (\d n_k \bd n_k)^{r-2p}   (\d n_m \d n_m)^p (\bd n_l \bd n_l)^p \ .     }
The 1-loop  renormalization 
matrix  has
 only the diagonal and  the next  off-diagonal 
line of entries, with the diagonal ones being proportional to \weg\ 
\eqn\ssaa{
 a_{1+p, 1+p} = (r-2 p) (r-1-2p) - 2 (\N-2) p \ , \ \ \ \  \ \ \ \ 
p=0, 1,2 , ...,  r/2  \ . }
The diagonal values are thus the eigenvalues, with largest one
for large $r$  
 being $a_{11}= r(r-1)$ (corresponding to  \zzz).  
 The associated eigen-operator \taps\  has all terms in the linear combination being present (i.e. all $c_p$ are non-vanishing). At the same time, 
the ``last'' operator is an eigen-operator by itself, 
\eqn\byby{
(\d n_k \d n_k \bd n_m \bd n_m)^{r/2}  \ \ : \ \ \ \  \ \ 
\ \ \   \g= 2-2r  -    r (\N-2) \bt   +  O(\bt^2) \ . }
Since $n_x^\el$  does mix with the scalar operators like
 $( \d n_k \bd n_k)  ^r$ or \taps\ (cf. \mop), 
it can be directly combined with them; using \opim,\zzz\ we get 
\eqn\zzi{ 
n^\el_x [ (\d n_k \bd n_k)^r+ ...] \ \ : \ \ \ \  \ 
 \g= 2-2r  - \ha   [\el(\el + \N-2)  -  2 r(r-1) ]\bt + 
O(\bt^2) \ . }
One finds  that due to the opposite signs of the two contributions
 the total  1-loop correction  to the anomalous dimension 
can be made small \pol\  for large $r$ by choosing $ \el \approx \sqrt 2  r$.
The analogous observation can be made \pol\ 
 for the corresponding \adn \  operator
 \eqn\zzii{ 
N^{-\D}_+ [ (\d N_M \bd N_M)^r+ ...] \ \ : \ \ \ \  \ 
 \g= 2-2r  + \ha   [\D(\D - \N+2)  -  2 r(r-1)] \bt + 
O(\bt^2) \ .  }
Provided a similar pattern persists at  higher loop orders
(i.e. the 2-loop correction is proportional to $\D^3 -  c r^3 + ...$ at 
large $\D$ and $r$, cf. \ssu,\zzz)  
one may  conjecture \pol\  that the 
dimension $\D(r)$ of  the corresponding 
high-level string vertex  operators should not blow up in 
the limit $\sql \to 0$.  
In the \adss string  context such  operator may  represent  
a  particular high-level scalar string mode with no  
extra global charges like spin or $S^5$ angular momentum.\foot{Such 
operators  may be dual to singlet  scalar SYM 
operators  of high dimension  like the ones considered in \mz.}

The 1-loop correction in \zzi\  has 
a very  similar  structure to the one of   the dimension of the 
``dilaton'' operator with  $S^5$ orbital momentum $J$ \tyu\ 
 in \margi. As we shall see below, an analogous 
 structure of the 1-loop anomalous dimension  is characteristic {\it  also }
 to the  vertex operators \mii,\enek\ 
  representing spinning string states  on leading Regge trajectory.

\subsec{\bf  Operators with angular momentum in $S^5$ }

We would like now to discuss renormalization 
 of the operator $( \d n_x \bd n_x)^{J/2}$,
which is expected to be  a part of the   vertex operator 
representing string state with angular momentum $J$ in the 5-sphere \jjj. 
  It is clear  from the above rules \boom, etc., 
that this operator may  mix with $\d n_k \d n_k$ 
and $\bd n_k \bd n_k$, but not with $\d n_k \bd n_k$.
On symmetry grounds,  the general  operator of the same  angular momentum $J$ 
 and dimension  $2-J$  (not containing  $\d n_k \bd n_k$ factors and higher 
derivatives) which is stable under the renormalization  
 is given by a  linear combination
of the following operators \mii\ 
(we shall assume that  $s=J/2$ is  even):
\eqn\lii{
O_s 
= \sum^{s/2}_{ p,q =0 } c_{pq}  M_{pq} \ , \ \ \ \ \
M_{pq} = V_{pq}+ V_{qp}  \ ,  \ \ \ \ \ \ \ \  s= J/2  \ ,   }
\eqn\derr{
V_{pq}=
 (n_x)^{2p+2q}  ( \d n_x)^{s-2p} (\bd n_x)^{s-2q}
(\d n_m \d n_m )^{p} (\bd n_k \bd n_k )^{q} 
 \ . }
We  have symmetrised the operators  over $p,q$ 
 so that there is the invariance  under $\del \to \bar \del$
 (or ``2-d reality'' property). 
The left and right sectors in $V_{pq}$  look 
completely independent so 
can be analysed separately.\foot{This would no longer be true for  
the operators with the same angular momentum and dimension
but containing extra factors of  $ \d n_m \bd n_m$.
In the flat space limit, such operators would  represent 
states on a subleading Regge trajectory.}

The analysis of renormalisation  of this operator  is similar to the
 case of \taps.
We find that under the 1-loop 
renormalization  the operators  $V_{pq}$ transform as follows
(here there is no summation under $p,q$) 
\eqn\sadd{
V_{pq} \to  V_{pq} + \bt I ( a_{pq}  V_{pq}   +  f_{p} V_{p+1,q}  +  f_{q} V_{p,q+1} )  \ , } 
where 
$$a_{pq}  =  - s (\N-1)  - (p+q) ( 2p+2q -1 )
 - 4(p+q)(s - p - q)  - (\N-2) (p+q)   - (s-2p) (s-2q)   \  $$
 \eqn\diaaa{
= 2 p^2 + 2 q^2 - (p+q) ( 2s + \N-3) -  s (s+ \N-1)  \ , }
\eqn\iaaa{
f_{p} = \ha (s-2p) (s-2p-1)  \ ,\   \ \ \ \ \ \ \   f_{q} = \ha (s-2q) (s-2q-1)  \ . }
Here the first $s (\N-1)$ term came from the $<n_k>$ in 
\asd\  and the last  one  -- from 
\dsa. The ``2-d real'' combination $M_{pq}= 
V_{pq} + V_{qp}$  transforms into 
a ``2-d real'' combination:
\eqn\akl{
M_{pq} \to  M_{pq} + \bt I ( a_{pq}  M_{pq}   +  f_{p} M_{p+1,q}  
+  f_{q} M_{p,q+1})   \  . } 
The anomalous dimension matrix is then  upper-triangular, and so 
 its eigenvalues are  equal to the diagonal entries $a_{pq}$.

The  minimal value of $|a_{pq}|$ is at $p=q= 0$, 
i.e.
\eqn\mini{a_{00} = - s( s+ \N-1) \ , \ \ \ \  }
 while  the maximal value is
 formally reached at  the extremum of the quadratic polynomial \diaaa, 
$p=q= \ha s +  { 1\ov 4} ( \N-3) $, i.e. at $ p=q= \ha s$ in the present case, 
\eqn\maxi{
a_{{s\ov 2}, {s\ov 2}}  =  -2s ( s + \N -2)  \ .  }
The associated  eigen-operators 
are particular linear combinations \lii.
The eigen-operator  for the the eigenvalue  $a_{pp}$ 
starts   with the term  in \lii\ which has the  coefficient $c_{pp}$.
In particular,  the operator $(\d n_x \bd n_x)^s$ is 
present only in the combination corresponding to the minimal 
eigenvalue \mini, i.e. in this case the eigen-operator 
contains all terms in \lii\ 
\eqn\pops{ O_s = c_{00} (\d n_x \bd n_x)^s + ... 
  + c_{{s\ov 2}, {s\ov 2}}
  (n_x)^{2s} (\d n_k \d n_k \bd n_m \bd n_m)^{s/2}  \ .  } 
Remarkably, in  contrast to what happens in flat space, 
in $S^{\N-1}$ (or $AdS_{\N-1}$) space an   operator 
with a factor $(\d n_x \bd n_x)^{J/2}$  representing a string state 
with an  ``intrinsic'' angular momentum   
may mix  with  an operator with a  factor 
$n_x^J$ which  may be interpreted as  representing   a state with 
``orbital''   component  of the angular momentum. 
At the same time, the ``last'' operator in the sum \lii\  or \pops\ 
does not mix with others --
it is by itself  the eigen-operator  for  the eigenvalue \maxi.

To construct a string vertex operator of canonical dimension $2-J$ 
 corresponding to a string state with angular momentum  $J$ 
in $S^5$ and energy $E=\D$ in $AdS_5$ we need  to multiply the above 
operators by $N_+^{-\Delta}$. This way we get  for the  case
 \mini\
($\N=6, \ s=J/2$)
\eqn\west{   N_+^{-\Delta} [ ( \d n_x   \bd n_x)^{J/2} + ...  ] \ \ :  }
\eqn\ress{
\g(\Delta, J)  = 2- J     +  
{ 1 \ov 2 \sqrt \l}[   \D (\D-4)  -  \ha  J  (J + 10 )  ] 
+  O( { 1 \ov (\sqrt \l )^2})  \ ,     }
and for the case \maxi\ 
\eqn\east{
 N_+^{-\Delta} n_x^J ( \d n_k \d n_k  \bd n_m \bd n_m)^{J/4} \  \ : 
 }
\eqn\kss{
\g(\Delta, J)  = 2-J    +  
{ 1 \ov 2 \sqrt \l}[   \D (\D-4)  -  J (J + 8)  ] 
+  O( { 1 \ov (\sqrt \l )^2}) \ .      }
Note that the coefficient $J(J+8)$ in \ress\   
is the sum of the 1-loop contribution
 $J(J+4)$ of $n_x^J$ in \opim,\margi\
and $ 4J$ of the operator \byby\ with $r= J/2$.\foot{In general,
the anomalous dimensions of particular operators 
should  be related to eigenvalues of generalized massive  Laplace
equation on a sphere. In particular, 
1-loop dimensions in \mini,\ress\ should correspond to 
special  symmetric tensor harmonics  on $S^{\N-1}$
 (cf. \hig).}

The 1-loop correction  in \kss\  has
 the same  structure as in 
 the dimension \margi\ of  the operator \tyu\ 
from the  ``massless''  level  carrying orbital  momentum $J$, 
i.e. 
 $ \g(\Delta, J)  =  
{ 1 \ov 2 \sqrt \l}[   \D (\D-4)  -   J  (J +4 )  ] 
+  O( { 1 \ov (\sqrt \l )^2})$.
At the same time, \ress\ has the same form as the dimension
 of the scalar $AdS_{\N-1}$ operator 
in \zzii,  which for the same level number  $2r=J$ 
as in \ress,\kss\ and $\N=6$ 
is given by  
$$ \g(\Delta, J)  =  2-J + 
{ 1 \ov 2 \sqrt \l}[   \D (\D-4)  -  \ha  J  (J - 2 )  ] 
+  O( { 1 \ov (\sqrt \l )^2})\ . $$
All of these operators share with \zzi\ the property \pol\ 
that the 1-loop anomalous dimension can be made small 
by balancing the $\D$-contribution against the $J$-contribution.
Thus imposing  marginality 
for large 
``classical'' values of  $J$ and $\D$  one finds that $\D= J + ...$ 
 for \margi,\kss\  and 
\eqn\stra{
\D= { 1 \ov \sqrt 2} J + ... }
 for \ress,\zzii\  
(assuming  that  higher order
 corrections do not spoil this conclusion, 
cf. \ssu,\dqg).

Expecting   that the operator in \west\ 
should be representing 
a folded rotating string state in $S^5$ it may seem strange to find
the proportionality coefficient
in \stra\ to be  ${ 1 \ov \sqrt 2} $ instead of 1, as 
implied \pred\ by the corresponding classical solution in \gkp.
There may be several  possible explanations for this discrepancy. 
First, it may be that higher-order corrections in \ress\
(that {\it are} important for ``classical'' values of 
$J$ and $\D$) first need to be summed up as in  \dqg\ 
and only  then one finds $\D= J + ... $ in the limit 
$q= {J\ov \sql} \gg 1$. 
Another possibility is  that additional
fermionic contributions that we ignored  above change the coefficient 
${ 1 \ov 2} $ in front of $J^2$ term \ress\ into  1, and thus 
change $1 \ov \sqrt{2} $ in \stra\ into 1. Then assuming that 
all higher order corrections \ssu\ will also have the form $\D^{n+1} 
- J^{n+1}$, we would then get the agreement with the semiclassical
prediction $\D= J + ...$ in \pred.\foot{Finding subleading
correction in $\D(J)$ in \pred\ would in any case 
 require 
summing contributions from all orders in $1 \ov \sql$
expansion.}

Finally, it could be that  in spite of the fact that 
the operator \west\  has the required  angular momentum and 
flat-space limit it is not the right vertex operator 
to be associated  with the semiclassical rotating string state
in $S^5$.  
An indication of  which vertex operator 
is an adequate one could  come from the discussion in section 4.2
of an  attempt  to reproduce 
the relation between $\D$ and $J$ using 
semiclassical path integral approach. 
For string rotating around north pole of $S^5$ one has \gkp:
$ds^2 = d \theta^2 + \sin^2 \theta d \vp^2 + \cos^2 \theta d
\Omega_3^2$, $t= \k \tau, \ \vp = \om \tau, \ \theta=\theta(\s),$ 
and  
$\theta'^2 = \k^2 - \om^2 \sin^2 \theta$.
For example, while   the operator in \east\ has  the 
 factor $n_x^J$ that would
provide a natural  source for the angle $\vp$ (cf. 
the discussion in section 2.2), the derivative part  of it 
is completely $SO(6)$ invariant (i.e. does not 
select a particular rotation plane) and thus it is unlikely to provide 
the required source for the angle $\theta$.

To attempt to  carry out   the semiclassical argument 
sketched in  section 4.2
 we would need to decide
on specific form of the  operator in \pops,\west. 
In particular, one could try  to use just the leading term in it
given  explicitly in \west.  One could try to justify this provided 
the coefficients in the linear combination \pops\ were small in 
the limit of large $s= J/2$. However, this is not the case according
to \sadd,\iaaa: the mixing is not suppressed in the large $s$ limit.
Indeed, fixing the coefficient of the first term in \pops\ to be 1, 
the coefficients of other terms  are proportional 
to products of off-diagonal values divided over products 
of differences 
 $a_{pp} - a_{00}= 2 p^2+q^2 - (p+q) (2s + \N-1) .$
 


\subsec{\bf  Operators with spin  in $AdS_5$ }
The analog of the operator \west\  that carries spin in
$AdS_{5}$  would be the one in \sss. Computing its dimension in 
a naive way, i.e. ignoring its mixing with other operators 
we would get 
\eqn\xst{   N_+^{-\Delta} [ ( \d N_x   \bd N_x)^{S/2} + ...  ]
\ \ :  }
\eqn\xss{
\g_{0} (\Delta, J)     =  2- S     +  
{ 1 \ov 2 \sqrt \l}[   \D (\D-4)  +  \ha  S  (S + 10 )  ] 
+  O( { 1 \ov (\sqrt \l )^2})  \ ,     }
where we used that in the AdS case one is to reverse 
the sign of coupling compared to \ress. 
In contrast to \ress, here both contributions to the 1-loop 
anomalous dimension term have the  same sign and thus 
can not  cancel each other.\foot{ At  small $S$ and large $\D$ 
we would still get from $\g=0$ the usual  linear Regge
trajectory relation, $\D^2  \approx   2 \sql (S-2) $.
To interpolate to the expected large $S$ behaviour \pred\
it may seem that  we   need first to sum up all higher-loop
contributions to $\g$.}

However, as  already mentioned above, the operator \sss\ 
mixes with other operators \enek\ under the
renormalization, and so  one should  first 
 diagonalize the corresponding 
matrix of anomalous dimensions. 

Instead of  addressing the question of renormalisation of
operators \xst\ directly 
in the $AdS_{\N-1} $ context, 
it is useful to follow the same notation as in the previous 
subsection and 
 solve  the equivalent problem in the  $S^{\N-1}$ 
sigma model.
 Namely, we shall consider the renormalization of
the  $S^{\N-1}$  operators  of the type 
\eqn\yxe{
n_y^\ell ( \d n_x   \bd n_x)^{s} + ... \ , \ \ \ \ \ \ \ 
n_x= n_1 + i n_2\ , \ \ n_y= n_3 + i n_4 \ , \ \  }
which are the  direct analogs of \xst. 
The relevant $AdS_5$ case will be obtained by setting
\eqn\spee{
\N=6\ , \ \ \ \ \ \ \el = -\D\ , \ \ \ \ \ \ \ \ \ 
s=S/2 \ , \ \ \ \ \ \ \ \ge \to - {1\ov \sql}\ . }
There will be two different types of mixings: 
(i) in view of \opp, i.e. $ <n_y,\del n_x> = - I n_x \del n_y$, 
we will need to include operators with $n_x$ factors  with no
derivatives 
and with  factors where  derivatives act on $n_y$;
(ii)  pairings of 
$\d n_x$ with $\del n_x$ or
$\del n_y$, etc., will generate as in \derr\ terms with extra 
scalar factors $\del n_k \del n_k$ and $\bd n_m \bd n_m$.
As a result, a generic   operator which has
 the same quantum numbers
and canonical dimension  as \yxe\ and which is an 
eigen-vector   of the
corresponding  anomalous dimension matrix will have 
the form (we shall assume that $\el \geq 2s$) 
\eqn\ene{
O_{\ell,s} = 
\sum_{u,v=0}^s \sum_{p=0}^{ s-u \ov 2} 
 \sum_{q=0}^{ s-v\ov 2} c_{uv pq}  M_{uvpq}   \ ,\ }
 \eqn\upp{ 
M_{uvpq} \equiv 
 n_y^{\ell-u-v} n_x^{u+v} (\d n_y)^u (\d n_x)^{s-u-2p} 
(\bd n_y)^v (\bd n_x)^{s-v -2q } 
(\d n_k \d n_k)^p (\bd n_m \bd n_m)^q     \  .}
 The ``last'' term in this sum is  the
 direct analog of 
\east\ and is an eigen-operator by itself:  
\eqn\ast{
 n_y^{\ell} n_x^{2s} ( \d n_k \d n_k  \bd n_m \bd n_m)^{s/2} \ 
 \ : 
 }
\eqn\ihss{
\g(\ell,s)  = 2-2s    - 
{ 1 \ov 2 } \ge  [ (\ell + 2s)^2  + (\ell + 4s) (\N-2)       ] 
+  O( \ge^2) \ .      }
To compute the  dimension \ihss\ we noted 
that according to \asd\ 
 pairings  in $n_x ^l n_y^{2s}$ give the contribution 
$-[  \ha \el(\el-1) + \ha 2s (2s-1)  + 2s \el ] 
 = -\ha (\el+2s) (\el+2s-1)$. 
Adding the  contribution of $<n_a>$ and
of the scalar operator \byby\   we  end up with \ihss.
Note that for the corresponding $AdS_5$ operator 
one is to use \spee, 
 and so for large $S$ and $\D$ 
the leading term in the 1-loop correction 
vanishes  if $\D=S$. 

To simplify the discussion let us now ignore all  mixings
with the scalar operators 
$(\d n_k \d n_k)^p (\bd n_m \bd n_m)^q $. They
 can be readily included as in 
 section  5.3 and that leads only to upper off-diagonal terms in
 the  anomalous dimension matrix, i.e. should  not 
 change the
 eigenvalues of the anomalous dimension matrix. 
 The $p=q=0$ term in  \ene\ is 
 \eqn\pne{
O_{\ell,s} = 
\sum_{u,v=0}^s c_{uv}  M_{uv}   \ ,\ }
 \eqn\lpp{ 
M_{uv } \equiv 
 n_y^{\ell-u-v} n_x^{u+v} (\d n_y)^u (\d n_x)^{s-u} 
(\bd n_y)^v (\bd n_x)^{s-v  }     \  .}
 Using the rules of section 5.1 we conclude that under the 1-loop 
 renormalization 
$$ M_{uv} \to \ \  ( 1 +   \bt  I  a_{uv} ) M_{uv}  $$ 
$$
 - \ \bt I  \bigg (   (\el-u-v) [ (s-u) M_{u+1,v} + (s-v) M_{u,v+1}] 
+  (u+v) [ u  M_{u-1,v} + v  M_{u,v-1}]  
$$\eqn\rree{
+ (s-u) v M_{u+1, v-1}  + u (s-v) M_{u-1, v+1}
\bigg)  \ ,  }
where $
u,v=0,..., s$  and the diagonal renormalization coefficients are
$$ 
a_{uv}= - \ha [   \el(\el+\N-2) + 2  s (\N-1)] 
-  (\el-u-v) (u+v) - (u+v) ( 2s - u -v) - u v  - (s-u)(s-v)  
$$ \eqn\aaqq{
= \  - \ha [   \el(\el+\N-2) + 2  s (s+ \N-1)] 
- (u+v) ( \el + s -2 u- 2v)   - 2 u v  \ . }
A naive guess is that for  large $\el,s$
 the ``upper off-diagonal''
term $ (\el-u-v) [ (s-u) M_{u+1,v} + (s-v) M_{u,v+1}]$
  in \rree\ will be dominating over the other off-diagonal 
  terms, and then the eigenvalues of anomalous 
  dimension matrix  will be  given by  the diagonal values
  \aaqq.
  For $u,v=0$  \aaqq\ becomes 
\eqn\jiu{  a_{00} = 
 - \ha    \el(\el+\N-2) -   s (s+ \N-1)  \ , }
 which  is related to the 1-loop coefficient in 
  \xss\ in  the special case \spee.
  For 
 $u=v=s$    we get  
\eqn\larr{
a_{ss}=  - \ha  \el(\el+\N-2) -  s (s+ \N-1)
- 2s\el  + 4 s^2    \ ,  }
so that $|a_{00} | < |a_{ss} |$ for $\el > 2s$.

 It is possible to find the eigenvalues of the 
 anomalous dimension  matrix in \rree\ following the  method 
 described in section  9 of \weg.\foot{I am grateful to F. Wegner 
 for an important clarification  of this method.}
 In general, the eigen-operators can be  characterised by  three
 Young tableaux: $Y_\d$ and $Y_{\bd}$ that describe 
 linear combinations  of permutations of the $2s$ derivatives 
 $\d$ and $2s$ derivatives $\bd$ in \lpp, respectively, 
 as well as by $Y_n$ that describes permutations of $\ell + 2s$ 
 factors $n_x$ and $n_y$. Then the  coefficients 
 in  eigenvalues of the anomalous dimension matrix 
  normalised as in 
 \aaqq\ are given by \weg\
 \eqn\taak{
 a= - \ha ( \el + 2s) ( \N-1)  + a_{\rm exchange} \ , }
 where the first term comes from the renormalization  (cf. \asd) 
   of the   $n_x$ and $n_y$ factors and 
 \eqn\exx{
  a_{\rm exchange} = \xi(Y_\d) + \xi(Y_{\bd} ) - \xi(Y_n) \ .}
  Here $\xi(Y)$ is an eigenvalue of the sum of transpositions 
  \eqn\jop{
  \xi(Y) = \ha \sum_{i=1,2,...}    k_i ( k_i - 2i + 1) \ , \ \ \ \ \ 
  Y= (k_1,k_2,...) \ , } 
  where $k_i$ are the  numbers of boxes in  the 
   rows of  a Young tableau $Y$. 
  Since \lpp\ contains only first powers of derivatives, 
  we get 
  \eqn\gett{
  Y_\d = (s,0,0,...) \ , \ \ \ \ \ \ \ Y_{\bd }= (s,0,0,...)
   \ , \ \ \ \
	  \xi(Y_\d) = \xi(Y_{\bd}) = \ha s (s-1)  \ .}
Given that we have only  two types of $n_k$-components,
 i.e. $n_x$ and $n_y$, 
$Y_n$ will have only two rows at most
 (we assume that $\ell \geq  2s$) 
\eqn\twoo{
Y^{(k)}_n= (\el+2s- k, k, 0, 0,...) \ , \ \ \ \  k=0,1,...,2s \ ,  }	  
\eqn\upp{
  \xi(Y^{(k)}_n) = \ha  ( \el +2s -k)  ( \el +2s -k - 1 )
  + \ha  k   ( k - 3 ) \ . } 
  Thus the eigenvalue coefficients \taak\  are given by 
  \eqn\taak{
 a^{(k)}
 = - \ha ( \el + 2s) ( \N-1)  +  s (s-1)    
 -   \ha  ( \el +2s -k)  ( \el +2s -k - 1 )
 - \ha  k   ( k - 3 )  \ , \ \ \ \ \    k=0,1,...,2s \ .  
  }
 $  a^{(k)}$  has a minimum at $k=0$ and a maximum at $k=2s$ 
  with values  (cf. \jiu,\larr) 
  \eqn\toak{
 a^{(0)}
 = - \ha  \el ( \el + \N-2 )  -  s ( s + \N-1 )  
  -  2 \el s  \ ,    }
  \eqn\tok{
 a^{(2s)}
 =      -   \ha   \el  ( \el  + \N - 2)  - s ( s + \N  - 3 )  \ .  }
  Written in $AdS_5$ notation \spee\  these expressions 
  become  (cf. \xss) 
   \eqn\toak{
 a^{(0)}
 = - \ha [  \D ( \D - 4)  + \ha   S ( S + 10  )  - 2  \D S  ] \ ,    }
  \eqn\tok{
 a^{(2s)}
 =   -   \ha  [  \D ( \D -4)  + \ha  S ( S + 6 ) ]  \ .  }
 Interestingly, in contrast to the naive dimension \xss\
 where the 1-loop term was positive for large $\D$ and $S$,
 here  
 $a^{(0)}$ can  be made to vanish.
 The corresponding  anomalous dimension 
 \eqn\wss{
\g (\Delta, J)     =  2- S     +  
{ 1 \ov 2 \sqrt \l}[   \D (\D-4)  +  \ha  S  (S + 10 )  - 2 \D S 
  ] 
+  O( { 1 \ov (\sqrt \l )^2})     }
 can thus   be made small for large $\D$ and $S$  provided 
 ($\D,S \gg \sql \gg 1$) 
 \eqn\zerr{ 
 \D(S)_{{S\ov \sql} \gg 1}  
 \approx ( 1 \pm { 1 \ov \sqrt 2} ) S  + \sqrt{2} \sql + O(1)   \ . }
 As a result,  for large $\D,S$ 
 one is able to construct a marginal vertex operator 
 with the  required quantum numbers.
 This is,  at least 
 qualitatively,  consistent with 
 the semiclassical prediction \pred. 
 To reproduce the same  proportionality 
 coefficient (i.e. 1) between $\D$ and $S$  as in \pred\
 one probably needs to sum up all relevant higher-order 
 $1\ov \sql$ corrections in \wss\ 
 (cf. \dqg).
 
Our main conclusion is that, as in the $S^5$ rotation case 
in section 5.3,  the vertex operators  that correspond to 
string states with  spin  in $AdS_5$  are given by linear
combinations of local operators like \pne. 
That seems to complicate possible 
rederivation of the  semiclassical prediction \pred\ 
for $\D(S)$ from 
the  string path integral expression 
for the 2-point correlator of the two 
vertex operators.

\bigskip 
\bigskip 

{\bf Acknowledgments}

\noindent
We are grateful for S. Frolov  for a collaboration 
at an initial stage  of this project. We would  like also 
to thank K. Zarembo for many important discussions.
 We also acknowledge  I. Klebanov, R. Metsaev   and
   F. Wegner  
 for discussions and very useful  explanations.
This  work of was supported in part
 by  the grants DOE DE-FG02-91ER40690,   PPARC SPG 00613,
INTAS  99-1590,  and by  the Royal Society Wolfson Award.

\vfill\eject
\listrefs
\end

\appendix{A}
{     Some basic relations       }

$z$ is the ``radial'' coordinate of
$AdS_5$ in the  Poincare  parametrization, i.e.
($m=0,1,2,3,$ \ $  p=1,...,6 $)
\eqn\asa{
(ds^2)_{_{AdS_5 \times S^5}}= {R^2 \ov z^2} (dx_\m dx_\m +
dz_p dz_p  ) \ , \ \ \ \
\ \ \ \ z^2 = z_p z_p
\ . }
In what  follows we shall often set $R=1$.

In  global coordinates
\eqn\add{
ds^2_{_{AdS_5}}
= - \cosh^2 \r \ dt^2 +  d\r^2 + \sinh^2\r \ d\Om_3 \ , }
$$
d\Om_3
= d \b_1^2 + \cos^2 \b_1 ( \ d\b_2^2 +
 \cos^2 \b_2  \ d \beta_3^2)   \ ,
 $$
while the  angle   $\vp$
of $S^5$  related to $z_p$ coordinates by
$dz_p dz_p = dz^2 + z^2 d\vp^2  + dz_n dz_n$.

In  general, the
 transformation between the Poincare and the global
 coordinates of $AdS_5$
can be  done as follows (we set the radius of $AdS_5$
 to be 1 and use the Minkowski signature):
\eqn\tran{
X_0={x_0\ov z} = \cosh \r \ \sin t \ , \ \ \
X_5= { 1 \ov 2 z} ( 1 + z^2 - x^2_0 + x_i^2) =  \cosh \r
\ \cos t \ , }
\eqn\med{
X_i= {x_i\ov z} = n_i \sinh \r  \ , \ \ \
X_4={ 1 \ov 2 z} ( -1 + z^2 - x^2_0 + x_i^2) =  n_4
\sinh\r \ ,
\ \ n^2_i + n^2_4 =1 \ ,   }
\eqn\sed{ \tan t = { 2 x_0 \ov 1 + z^2  - x^2_0 + x_i^2 }
  \ , \ \ \ \ \ \ \ \
z^{-1} = \cosh \r \ \cos t - n_4 \sinh \r  \ . }
Here $X_0,X_i,X_4$ ($i=1,2,3$)
are the coordinates of  $R^{2,4}$:
the $AdS_5$ metric is induced from the flat $R^{2,4}$ one
by the  embedding
$$X_0^2 + X_5^2 - X_i^2 - X^2_4 =1 \ . $$
The unit vector
 $n_k$ ($k=1,2,3,4$), \     $n^2_i + n^4_4=1$, \
 parametrizes the 3-sphere: \ $dn_k dn_k = d\Omega_3$.
The obvious point-like solution in global
coordinates \add, i.e. $t= \om \tau, \ \r=0, \  \vp= \om
\tau $
 (with all other angles being trivial),
then becomes  equivalent  to
\eqn\zes{  t= \om \tau\ , \ \ \ \  \vp = \om \tau  \ ,\ \
\ \ \  \
 z=  { 1 \ov \cos \om \tau}   \ . }
Here $\vp$ is, in fact, the  angle of large  circle of
$S^5$
and $t$ is the global time  coordinate of $AdS_5$.

The energy in global coordinates expressed in terms of
the   Poincare coordinates
$x_\m,z_p$ in \asa\
as functions of  both  $\tau$ and $\s$  is
$$E= \sql \E =  \sql\int {d \s\ov 2\pi}  \E_d =
    \sql\int {d \s\ov 2\pi}  { \rm cosh}^2\r\  \dot t
$$
\eqn\eee{
 =   \ha   \sql\int {d \s\ov 2\pi}  \big[ (1 + z^2  + x^2
 ) \PP_0 - 2 x_0 \D \big
] \ ,
}
\eqn\tyi{    \PP_0 =   {1 \ov z^2} \dot x_0 \ ,  \ \ \ \
\  \ \ \ \
\DD=   {1 \ov 2z^2} { \del \ov \del \tau} ( z^2  + x^2 )
 \ , \ \ \ \ \ \
  x^2= - x_0^2 + x_i^2\  .  }

For the solutions  discussed below  one has the following
relation
 \eqn\spe{
 z^2 - x^2_0 + x^2_i = 1       \ . }

Next,  let us    summarize the (conformal-gauge) form of
the
simplest classical string solutions  in \adss
written  in global  and Poincare  coordinates
of $AdS_5$
($G$ will stand for the  global and $P$ for the  Poincare
coordinate form).
As was already  discussed above, for  the
point-like string  rotating in $R^6$ (or, equivalently,
boosted
along a big circle in $S^5$)
\eqn\poi{
G: \ \ \ \   t= \n \t \ , \ \ \ \ \ \ \  \vp = \n \t  \ ,
}
\eqn\poo{ P: \ \ \   x_0 = \tan t \ , \ \ \  z= { 1 \ov
\cos t}  \ , \ \ \ \ \
\vp =  t = \n \t \ . }
The spinning string in $AdS_5$ is described by (the angle
$\p$ is $\beta_3$ in \add) \rf{\dev,\gkp}
\eqn\spin{
G: \ \ \ \   t= \k \t \ , \ \ \ \p = w \t \ , \ \ \r=
\r(\s) \ , \ \
 \ \  \r'^2 = \k^2 \cosh^2 \r - w^2 \sinh^2 \r   \ , }
or, equivalently, by
\eqn\pin{  P: \ \ \
 x_0 = \tan t \ , \ \ \  z= { 1 \ov \cos t
\ \cosh \r} \ , \ \ \
 x_1 = r \cos \p \ , \ \  x_2 = r \sin \p \ ,
 \  \  r\equiv  { \tanh \r\ov  \cos t} \ , }
where $t=t(\t)$, \ $\p=\p(\tau)$, and $\r=\r(\s)$ are
given by
\spin.
Here  $\r$ changes from 0 to its maximal value
$\r_0=$Arctanh${\k\ov w} $
and
the parameter $w$  is a function of $\k$.
In Poincare coordinates the string moves towards
the  horizon (center of AdS), rotating and stretching.
More general solution can be obtained  by combining the
rotation in $AdS_5$
with the  boost in $S^5$ \frt.

The form of a
 classical  solution cannot  depend on the value of the
 string tension,  i.e. on $\sql$, which appears as  a
 factor in front
the string  action
$$I= {\sql \ov 4 \pi} \int d^2 \xi \
G_{MN} (x) \del_a x^M \del^a  x^N \ .$$
 Thus the classical
energy
can be written as  $E= \sql \E(\cN)$, where $\cN$ stands
for all
constant parameters
that enter the classical solution. These parameters
should be
 fixed in the standard sigma model loop ($1 \ov \sql $)
 expansion.
However,  some of them may be quantized in the full
quantum theory,
i.e. $\sql \cN = \rN$=integer
(being related to canonical momenta the  quantized
parameters should contain
a factor of string tension).
For example, the  global charges  like  the
$S^5$ and $AdS_5$ angular momentum  components
$J= \sql \nu, \ \ S= \sql \S (\k)$, etc.,   will  take
integer values.

\vfill\eject
\listrefs
\end

In  AdS we do not have usual S-matrix; proper string vertex operators
are localised at the boundary (in $x_\m$) and have specific dependence
on radial direction  $z$.  To develop  semiclassical method
we may consider two small  string states or loops (cf. Wilson loops)
at the boundary  and look for  classical  string solutions that connect
them  in \adss.  That will determine a semiclassical
 expression for the string field propagator.  In the limit when
 boundary contours are point-like and simply delta-functions
 we get essentially semiclassics for KG operator in AdS \yon.
 Then euclidean trajectory gives us 2-point function for the $\D=J$
 operator. This is pure field theory
 in AdS space  --  no string theory used  so far.

A more non-trivial case is  rotating string:
suppose we consider  two  sting states  corresponding
to rotation in $x_1,x_2$  and separated in $x_3$ at the boundary.
Naive guess is that to do that all we need
is to consider Euclidean (both spacetime and world sheet)
continuation of the rotating string solution
written in Poincare coordinates  and to evaluate the string action on
it. We will presumably need a  cutoff $z=\epsilon$
plus may be some other  regularization
(cf. the Minkowski w-sheet was singular at the turning points \frt).

This is a heuristic guess, but it is probably correct
as we know that all works simply on the cylinder-- we get the right
dimension  from the classical energy. So presumably
on the 2-sphere the role  of vertex operators is only to
insert proper boundary conditions in mapping 2-sphere to the cylinder,
detail pre-exponential form of the vertex operator is not that
important.\foot{One may hope that just
evaluating the euclidean string action
on the euclidean solution  with spin $S$
(and which  starts at
point $x_3$ and returns to point $x_3'$)  will give the required result
for the correlator. This will  go beyond the previous example with
large $J$  which was usual point-particle case.}

Semiclassics here will be important  simply  because we
will be ignoring fluctuations as in the Wilson loop examples
plus  we  will have classical (i.e. large $\sim \sql$)
value  of the  spin $S$.
We will need single fold to minimize the action.


Summary of  attempts to understand  relation between space-time conformal
dimension $\Delta$ and  spin
 $S$  for large $S$ from  the
\adss string theory path integral representation
for correlators of string  vertex operators.
 More precisely, we shall start
 from the condition of marginality of the corresponding vertex
operator
following the suggestion  of Polyakov in the large
$J$ case.
The motivation is to  understand better the structure of
vertex operators, at least in the semiclassical limit
(when fermions can be ignored)
 and thus potentially the  relation between the
 correlators of local gauge-theory
operators  and  local string vertex operators
on the two sides of the duality.

Main points:
we first  determine the
vertex operator corresponding to the  string modes on
leading Regge  trajectory and try to compute its
dimension as function of spin  in semiclassical approximation in
path integral  following the same
 idea as suggested by Polyakov
in the case of the vertex operator for the
point-like (graviton)  mode carrying large momentum in $S^5$.
We clarify the meaning  of complex trajectory in path integral.
Possible application is determination
of $\a'\sim \l^{-1/2} $-corrected part  of the Klein-Gordon equation
for the space-time field representing the
massive higher spin string mode.
Taking $\l\to 0$ one may be able to see how
this mode becomes massless ....
Our calculation may have generalizations to other
cases, in particular, to near-conformal ones.

It is non-trivial that this  non-local classical string solution
 is
actually -- as in the flat space -- described by a local vertex
operator.  Thus AdS/CFT  should extend to string modes in an
expected or  ``usual''  way -- as relation between local
gauge theory operators and local string vertex operators.
This is actually quite non-trivial, if possible at all.